\documentclass[a4paper,11pt]{article}

\usepackage[OT1]{fontenc}
\usepackage{amsthm,amsmath}
\usepackage{natbib}
\usepackage[colorlinks,citecolor=blue,urlcolor=blue]{hyperref}
\usepackage{natbib}
\usepackage{amssymb,mathrsfs}
\usepackage{graphicx,xcolor}
\usepackage{subfig}
\usepackage[margin=25mm]{geometry}

\newcommand{\matr}{}
\newcommand{\vect}{\boldsymbol}
\newcommand{\gam}{\mathrm{Ga}}
\newcommand{\expo}{\mathrm{Exp}}
\newcommand{\norm}{\mathrm{N}}
\newcommand{\bet}{\mathrm{Beta}}
\newcommand{\dir}{\mathscr{D}}

\providecommand{\keywords}[1]
{
  \small
  \textbf{\textit{Keywords---}} #1
}

\title{Incorporating compositional heterogeneity into Lie Markov models for phylogenetic inference}
\author{Naomi E.~Hannaford$^1$, Sarah E.~Heaps$^{1,*}$, Tom M.~W.~Nye$^1$,\\ Tom A.~Williams$^2$ and T.~Martin Embley$^3$}
\date{
\small $^1$School of Mathematics, Statistics and Physics, Newcastle University, Newcastle upon Tyne, U.K.\\
\small $^2$School of Biological Sciences, University of Bristol, Bristol, U.K.\\
\small $^3$Institute for Cell and Molecular Biosciences, Newcastle University, Newcastle upon Tyne, U.K.\\
\small $^*$Corresponding author: \texttt{sarah.heaps@ncl.ac.uk}}

\begin{document}

\maketitle

\begin{abstract}
Phylogenetics uses alignments of molecular sequence data to learn about evolutionary trees. Substitutions in sequences are modelled through a continuous-time Markov process, characterised by an instantaneous rate matrix, which standard models assume is time-reversible and stationary. These assumptions are biologically questionable and induce a likelihood function which is invariant to a tree's root position. This hampers inference because a tree's biological interpretation depends critically on where it is rooted. Relaxing both assumptions, we introduce a model whose likelihood can distinguish between rooted trees. The model is non-stationary, with step changes in the instantaneous rate matrix at each speciation event. Exploiting recent theoretical work, each rate matrix belongs to a non-reversible family of Lie Markov models. These models are closed under matrix multiplication, so our extension offers the conceptually appealing property that a tree and all its sub-trees could have arisen from the same family of non-stationary models.

We adopt a Bayesian approach, describe an MCMC algorithm for posterior inference and provide software. The biological insight that our model can provide is illustrated through an analysis in which non-reversible but stationary, and non-stationary but reversible models cannot identify a plausible root.

\end{abstract}





\section{Introduction}
\label{s:intro}
The goal of phylogenetic analysis is to learn about the evolutionary relationships among a collection of species, typically using molecular sequence data. In this paper, we focus on sequences of DNA, with its four-character alphabet $\Omega = \{ \texttt{A}, \texttt{G}, \texttt{C}, \texttt{T} \}$. The evolutionary relationships are expressed graphically in the form of a bifurcating tree called a phylogeny whose external nodes, or ``leaves'', represent the species of interest, with ancestral species (speciation events) represented by internal nodes. In particular, the root of the tree is identified with the most recent common ancestor (MRCA) of all the species at the leaves. Determining the branch on which this root lies is fundamental to the biological interpretation of the tree because it fixes the direction of ancestry and provides a tool for tracing the evolution of important traits along the tree.

During reproduction, when an organism passes a copy of its DNA to its offspring, point mutations can occur in molecular sequences. When a point mutation becomes fixed in a population it is referred to as a \emph{substitution}. In statistical phylogenetics, substitutions are generally modelled using continuous time Markov processes (CTMPs). To this end, denote by $Y(t) \in \Omega$ the nucleotide at a particular site in a DNA sequence at time $t \ge 0$. Standard models of sequence evolution generally assume that the process $\{ Y(t): t \ge 0 \}$ is \label{pg:rev1_p2}stationary and time-reversible, meaning the joint probability of $\{Y(t) = i, Y(t + s) = j \}$ is the same as that of $\{Y(t - s) = j, Y(t) = i \}$ for all $0 \le s \le t$. These assumptions offer mathematical convenience, allowing the instantaneous rate matrix characterising the CTMP to be factorised as the product of a symmetric matrix of \emph{exchangeability} parameters, indicating a general propensity for change between pairs of nucleotides, and a diagonal matrix of stationary probabilities. Unfortunately, conditional on any particular (unrooted) tree, a mathematical consequence is that the likelihood does not depend on the position of the root. Traditionally, other methods have therefore been used to root evolutionary trees. The most common strategy, called \emph{outgroup rooting}, requires inclusion of data from a set of taxa (the \emph{outgroup}) which are known to have evolved outside the subtree of interest (the \emph{ingroup}). The root can then be placed on the branch connecting the outgroup to the rest of the tree. Unfortunately, this approach can be problematic when the outgroup is only very distantly related to the ingroup. In such cases, model assumptions become increasingly dubious and the ensuing model misspecification can result in the identification of spurious relationships with the outgroup.

An alternative strategy for rooting, which has received relatively little attention in the phylogenetic literature, is to take a model-based approach and draw inference from a likelihood that depends on the position of the root. In standard models, likelihood invariance to the root position arises as a consequence of the assumptions of stationarity and time-reversibility. Although mathematically convenient, these assumptions typically do not hold up under the scrutiny of biological examination. For example, in a stationary process, time-reversibility implies that the direction of time is unimportant. Yet the mechanism by which substitutions in DNA occur is very complex, comprising processes of point mutation and fixation, and encompassing the effects of selection, and so on. Whilst most physical processes are time-reversible, when these sub-processes combine to produce substitutions in DNA sequences, their complex interaction makes time-reversibility highly questionable. Correspondingly, analysis of biological data often provides evidence to rebut the reversibility assumption \citep[][]{squartini}. Similarly, there are various biological theories which are discordant with a stationary evolutionary process, for example mutational biases in DNA replication enzymes are thought to differ across the domains of life \citep[][]{LA08}. The assumption is also easy to challenge empirically. In any particular analysis, if the taxa had evolved according to a stationary process, one would expect the sequence composition of each taxon to be approximately the same. \label{pg:rev1p3}Yet, in analyses of deep phylogenies, the GC-content (the total proportion of \texttt{G} and \texttt{C} nucleotides) of 16S ribosomal RNA, the most widely used gene in phylogenetic analysis, varies from 45\% to 74\% across the diversity of sampled Bacteria, Archaea and eukaryotes \citep[][]{cox}. Similarly, heterogeneity in sequence composition has also been observed in much more recent species radiations, for example, placental mammals \citep[][]{morgan2013}, marsupials \citep[][]{phillips2006}, birds \citep[][]{braun2002} and paraneopteran insects \citep[][]{li2015}. As a consequence, in addition to facilitating root inference, models that relax the restrictive assumptions of stationarity and / or reversibility also provide opportunities to incorporate further biological realism.

In this paper we develop two non-stationary \label{pg:rev2_p4}models for DNA evolution. Substitutions along each branch of the tree are driven by a different instantaneous rate matrix, with non-reversible structure, allowing step changes in the theoretical stationary distribution over time. \label{pg:rev1p1}The resulting models capture the lineage-specific drift in sequence composition that is often hypothesised biologically and are shown empirically to allow inference on the position of the root. Leveraging recent theoretical work, the branch-specific rate matrices are Lie Markov models \citep[][]{sumner2012a} which offer the property of closure under matrix multiplication. Such models are conceptually appealing because they provide a framework in which a tree and all its sub-trees could have arisen from the same family of non-stationary models.

The remainder of this paper is structured as follows. In Section~\ref{sec:std_models} we provide an overview of existing substitution models for phylogenetic inference. Section~\ref{sec:nonhomo_lie} then introduces a pair of non-reversible Lie Markov models, and our non-homogeneous, non-stationary extension. In Section~\ref{sec:prior} we describe the priors for the parameters in our two non-homogeneous models. Section~\ref{sec:posterior} outlines the structure of the posterior distribution and our Markov chain Monte Carlo (MCMC) algorithm for computational inference. In Section~\ref{sec:simulations}, we use simulation experiments to verify, empirically, that the root position in our non-homogeneous models is identifiable under the likelihood and investigate rooting performance under a variety of conditions. Section~\ref{sec:application} then considers an application to a data set for which root inference has previously been challenging. Finally, we summarise in Section~\ref{sec:discussion}.

\section{\label{sec:std_models}Models of DNA evolution}

Denote by $\tau$ a phylogeny, with branch lengths $\vect{\ell}$, representing the evolutionary relationships among a collection of $n$ taxa (species). Consider the nucleotide (letter) at a single genomic site in the MRCA at the root of the tree. Over time, substitutions may have accumulated at that site such that the corresponding sites in the $n$ taxa at the leaves of the tree are occupied by different nucleotides. The ensuing assignment of \texttt{A}, \texttt{G}, \texttt{C} or \texttt{T} to each taxon is referred to as the \emph{DNA character} at that site. There are clearly $4^n$ possible DNA characters for a phylogeny on $n$ species and we denote this set by $\Omega^n$.

Suppose we have an \emph{alignment} of data $\matr{y} = (y_{ij})$, with $n$ rows, representing the $n$ species at the leaves of the tree, and $m$ columns, representing genomic sites in the MRCA. We assume that the molecular sequences of each of the $n$ taxa have been \emph{aligned} such that the columns can be regarded as observations of a DNA character. 

\subsection{\label{subsec:standard_mods}Standard models}

Denote by $Y(t) \in \Omega$ the nucleotide at time $t$ at a single genomic site and consider evolution along a single branch of the phylogeny $\tau$. Most phylogenetic models assume that substitutions can be modelled using a CTMP, characterised by an instantaneous rate matrix $\matr{Q} = (q_{uv})$. Over some interval of time, represented by $\ell$, the transition probabilities between pairs of nucleotides are obtained by taking the matrix exponential $\matr{P}(\ell) = \exp(- \ell \matr{Q}')$ where $\matr{Q}'=\matr{Q} / (- \sum_u q_{uu} \pi_u)$ and $\vect{\pi} \in \mathscr{S}_4$, \label{pg:rev2_p5}$\mathscr{S}_K = \{ \vect{x} = (x_1, \ldots, x_K): \; x_i \ge 0 \forall i, \; \sum x_i = 1\}$, is the stationary distribution of the process, satisfying $\vect{\pi} \matr{Q} = \vect{0}^T$. This rescaling of the rate matrix allows the branch length $\ell$ to be interpreted as the expected number of substitutions per site. The $(u,v)$-element in $\matr{P}(\ell)$ is the probability of transitioning from nucleotide $u$ to nucleotide $v$ along a branch of length $\ell$, $p_{uv}(\ell) = \Pr\{ Y(\ell) = v | Y(0) = u \}$ for any $u,v \in \Omega$.

Standard phylogenetic models are typically based on three assumptions: \emph{(i) homogeneity} -- a single instantaneous rate matrix characterises the evolutionary process along every branch of the tree; \emph{(ii) stationarity} -- the CTMP is in its stationary distribution $\vect{\pi}$ and so $\vect{\pi}$ is also the distribution at the root; \emph{(iii) reversibility} -- the CTMP is time-reversible, that is $\pi_u p_{uv}(\ell) = \pi_v p_{vu}(\ell)$ for all $u, v \in \Omega$. The instantaneous rate matrix $\matr{Q}$ of a homogeneous, reversible process can be factorised as $\matr{Q} = \matr{R} \matr{\Pi}$ where $\matr{\Pi} = \text{diag}(\vect{\pi})$ is the diagonal matrix of stationary probabilities and $\matr{R}=(r_{ij})$ is the symmetric matrix of exchangeability parameters with $r_{ij} = r_{ji} \ge 0$ for $i \ne j$. We refer to a rate matrix as reversible if it permits a factorisation of this form.

In the class of reversible rate matrices, the most general is that of the general time reversible (GTR) model \citep{tavare}, with six distinct exchangeability parameters. Other commonly used models are then derived as special cases. For example, ordering the nucleotides as \texttt{A}, \texttt{G}, \texttt{C}, \texttt{T}, the HKY85 model \citep{hasegawa} is a special case, whose rate matrix $\matr{Q}$ is given in Figure~\ref{fig:Q1}.
In this model, the reduction in the number of exchangeability parameters from six to two is biologically motivated, allowing transitions (substitutions between purines -- \texttt{A} and \texttt{G} -- and between pyrimidines -- \texttt{C} and \texttt{T}) to occur at a different rate to transversions (substitutions between a pyrimidine and a purine).

To prevent arbitrary rescaling of the rate matrix $\matr{Q}$ in the transition matrix $\matr{P}(\ell) = \exp(-\ell \matr{Q}')$, an identifiability constraint is typically imposed. For reversible rate matrices, this often entails setting one exchangeability parameter equal to one \citep[][]{ZH04} so that the others can be interpreted as relative propensities for change. For example, fixing $\lambda=1$ in the HKY85 model in Figure~\ref{fig:Q1}, $\kappa$ is interpreted as the transition-transversion rate ratio. 

As explained in Section~\ref{s:intro}, the assumptions of stationarity and reversibility in a homogeneous model are often not justifiable from a biological perspective. Moreover, they come at an inferential cost, giving rise to likelihood functions that are invariant to the position of the root of the tree. As such, these models can only be used to infer unrooted trees, which depict the branching pattern of speciation events, without associating direction to the branches of the tree. Models which relax one or both assumptions can therefore offer more biological credibility, whilst also providing a likelihood function that is informative about the direction of time. We explore existing models of this type in the section which follows.

\subsection{\label{ss:rootinf}Models facilitating root inference}

Motivated by the rooting problem, \citet{huelsenbeck2002} and \citet{cherlin17} investigated stationary but non-reversible substitution models in a Bayesian framework. In each case, the model was based on an instantaneous rate matrix which was structurally unconstrained, representing the so-called \emph{general Markov model} of DNA evolution \citep[][]{barryHartigan}. Not surprisingly, simulation experiments and application to biological data sets suggested that these stationary, but non-reversible models can produce sensible root inferences when the model assumptions are clearly supported. However, root inference was found to be very sensitive to model misspecification, especially violation of the assumption of stationarity \citep[][]{williams15}. This limits the utility of such models in application to data sets of biological interest, where it is common to see variation in sequence composition across taxa due to lineage-specific compositional change.

\label{pg:rev1p1b}Most models which allow root inference are \emph{non-homogeneous}, which means that the process cannot be characterised by a single instantaneous rate matrix. Instead matrices from a countable set $\{ \matr{Q}_1, \matr{Q}_2, \ldots \}$ apply to different parts of the tree. In general, the $\matr{Q}_b$ all belong to the same family of rate matrices. For example, \citet{kaehler2017} considers a model in which the $\matr{Q}_b$ are all strand-symmetric, meaning the rate of substitution between a pair of nucleotides is the same as that between their Watson-Crick base pair complements. For this special class of models, a mathematical proof is provided which verifies that the root position can be identified from the likelihood. However, inferential methodology to fit the model to data has not yet been developed. More often in the literature, the $\matr{Q}_b$ all belong to a family of \emph{reversible} rate matrices, such as HKY85 or GTR, so that $\matr{Q}_b = \matr{R}_b \matr{\Pi}_b$ for $b=1,2,\ldots$. \label{pg:rev2_p4b}We refer to such a non-homogeneous process as \emph{locally} reversible. The resulting models are generally \emph{non-stationary}, with $\matr{\Pi}_b \ne \matr{\Pi}_{b'}$ for $b \ne b'$, and hence allow step changes in the theoretical stationary distribution, sometimes termed the \emph{composition vector}, across the tree. Some also allow variation in the exchangeability parameters \citep[][]{dutheilBoussau08}, although these are often constant, with $\matr{R}_b = \matr{R}$ for all $b$. For example, \citet{yang1995} investigated a model in which the exchangeability parameters were constant over the tree, but the composition vectors in the $\matr{\Pi}_b$ varied from branch to branch. \citet{heaps} investigated a similar model in a Bayesian framework, with a prior that assumed positive correlation among the set of composition vectors, thereby allowing information to be shared between branches. Other approaches intended to reduce the variance of parameter estimates in complex models of this form have been largely based on the idea of dimension reduction. For instance, \citet{foster} considered a mixture model in which the $B$ branches of the tree were allocated to one of $K \ll B$ mixture components, with branches in the same component sharing a composition vector. Similarly, \citet{blanquart} introduced a model in which the step changes in the theoretical stationary distribution occurred according to a Poisson process, independently of the speciation events which determine the tree's branching structure. The main difficulty with these mixture-type models is that the dimension of the problem -- determined by the number of mixture components in the former case and the number of ``break--points'' in the latter -- are not known \textit{a priori}, which substantially complicates computational inference. 

Lie Markov models for DNA evolution have the property of closure under matrix multiplication \citep{sumner2012a}.  Let $P_1$ and $P_2$ be transition matrices obtained by taking the matrix exponential of two rate matrices from a family of Markov models $\mathcal{M}$.  $\mathcal{M}$ is multiplicatively closed if and only if for all such $P_1,P_2$, the product $P_1 P_2$ is obtainable as the matrix exponential of another rate matrix from the family $\mathcal{M}$. \citet{woodhams} defined a hierarchy of Lie Markov models capable of distinguishing pairs of nucleotides, such as purines and pyrimidines. The general Markov model represents the most complex family of Lie Markov models. \label{pg:rev1_p4}It has twelve parameters and so eleven degrees of freedom after imposing an identifiability constraint to fix its scale; see Section~\ref{subsec:standard_mods}. However, all other Lie Markov models can be represented by ten parameters, and hence nine degrees of freedom, or fewer. Some are non-reversible and, like families of reversible rate matrices, have a biologically interpretable structure. It is therefore possible to combine the ideas from this section and build parsimonious models that are non-stationary, non-homogeneous and locally non-reversible by using an appropriate set of rate matrices from a family of Lie Markov models. This is the focus of Section~\ref{sec:nonhomo_lie}.

\subsection{Assumptions across sites}

The previous sections have described substitution models for evolution at a single genomic site. In order to extend this to a joint model for the whole alignment, sites are generally assumed to evolve independently, but with their own rates $\gamma_i$, $i=1,\ldots,m$, which scale the normalised rate matrix $\matr{Q}'$ linearly. Biologically, this reflects the idea that rates of evolution vary according to functional or structural pressures acting at a site: important sites are subject to higher selective constraints and hence evolve more slowly. \label{pg:rev1_p5}These site-specific parameters $\gamma_i$ are modelled as multiplicative random effects, $\gamma_i | \phi \sim \gam(\phi, \phi)$ for $i=1,\ldots,m$, where the common shape and rate $\phi$ give the distribution a unit mean. The value of $\phi$ controls the manner and extent to which evolutionary rates vary across sites. 

During model-fitting, discretising the continuous gamma distribution allows intermediate likelihood calculations to be cached, which substantially speeds up computation. Therefore, in keeping with standard practice in the phylogenetic literature, we adopt a discrete approximation to the gamma distribution with four rate classes \citep[][]{yang1994}. Under this distribution, the rate $\gamma_i$ at site $i$ is equal to $r_k(\phi)$, $k=1,2,3,4$, with probability $p_k = 1/4$ and $r_k(\phi)$ taken as the $(k - 0.5)/4$ quantile in the $\gam(\phi, \phi)$ distribution.

\section{\label{sec:nonhomo_lie}Non-homogeneous Lie Markov models}

In this section we begin by describing two families of Lie Markov models which, in the terminology of \citet{woodhams}, are referred to as the RY5.6b and RY8.8a Lie Markov families. In each case, we derive a new parameterisation of the underpinning rate matrix $\matr{Q}$, and the relationship between the new parameters and the theoretical stationary probabilities $\vect{\pi}$. We then introduce two non-homogeneous models which we construct by allowing the RY5.6b or RY8.8a rate matrix to vary from branch to branch.

\begin{figure}[!t]
\subfloat[\label{fig:Q1}]{
$
\begin{pmatrix}
* & \kappa \pi_2 & \lambda \pi_3 & \lambda \pi_4 \\
\kappa \pi_1 & * & \lambda \pi_3& \lambda \pi_4\\
\lambda \pi_1 & \lambda \pi_2 & *  & \kappa \pi_4\\
\lambda \pi_1 & \lambda \pi_2 & \kappa \pi_3 & * \\
\end{pmatrix}
$
}
\hfill
\subfloat[\label{fig:Q2}]{
$
\begin{pmatrix}
* & \alpha + \rho_2 & \beta + \rho_3 & \beta + \rho_4 \\
\alpha + \rho_1 & * & \beta+ \rho_3& \beta + \rho_4\\
\beta+\rho_1 & \beta + \rho_2 & *  & \alpha + \rho_4\\
\beta + \rho_1 & \beta + \rho_2 & \alpha + \rho_3 & * \\
\end{pmatrix}
$
}
\hfill
\subfloat[\label{fig:Q3}]{
$
\begin{pmatrix}
*              &\rho_2         &\tilde{\rho}_7 &\tilde{\rho}_8\\
\rho_1         &*              &\tilde{\rho}_7 &\tilde{\rho}_8\\
\tilde{\rho}_5 &\tilde{\rho}_6 &*              &\rho_4\\
\tilde{\rho}_5 &\tilde{\rho}_6 &\rho_3         &* 
\end{pmatrix}
$
}
\caption{Instantaneous rate matrices $\matr{Q}$ for the \protect\subref{fig:Q1} HKY85; \protect\subref{fig:Q2} RY5.6b; \protect\subref{fig:Q3} RY8.8a models. The values of * ensure that the rows sum to zero. The nucleotides are ordered \texttt{A}, \texttt{G}, \texttt{C}, \texttt{T}.}\label{fig:Q}
\end{figure}

\subsection{The RY5.6b model}

Motivated by its simplicity and similarity to the widely used HKY85 model in Figure~\ref{fig:Q1}, the first Lie Markov model we consider is the (non-reversible) RY5.6b model. Following the formulation presented in \citet{woodhams}, its rate matrix $\matr{Q}$ can be represented as in Figure~\ref{fig:Q2}
where $\alpha, \beta, \rho_1, \rho_2, \rho_3, \rho_4 \geq 0$. As indicated by the prefix of its name, the model has the symmetry condition of purine-pyrimidine (RY) pairing, with rates of change for transversions sharing a parameter and rates of change for transitions sharing a different parameter. However, the six parameters are plainly not identifiable since we can replace $\alpha$ and $\beta$ with $\alpha + \delta$ and $\beta + \delta$, and $\rho_i$ with $\rho_i - \delta$ for $i=1,2,3,4$, and obtain exactly the same rate matrix. In the reversible HKY85 case, the off-diagonal elements in each column of the rate matrix share a stationary probability $\pi_i$ with $\vect{\pi} \in \mathscr{S}_4$. In the RY5.6b model, they each share a parameter $\rho_i$. By choosing the analogous constraint, $\vect{\rho} = (\rho_1, \rho_2, \rho_3, \rho_4) \in \mathscr{S}_4$, we can eliminate the parameter redundancy. We note that the 5 and 6 in the name of the RY5.6b model arise from it being a five-dimensional model whose rate matrices form a polyhedral cone with six rays \label{pg:rev2_p6}\citep[][]{fernsand2015}. These allow it to be expressed through six non-negative parameters.

Although the simplex constraint removes the additive identifiability issue, the overall scale of the rate matrix is still arbitrary since it appears only in its normalised form, $\matr{Q}' = \matr{Q} / (- \sum_u q_{uu} \pi_u)$, in the transition matrix. To resolve this problem, it is convenient to fix the scale of the rate matrix by constraining its trace to be equal to -7 as this limits the support of the remaining parameters so that $(\alpha, 2 \beta) \in \mathscr{S}_2$ or, equivalently, $\alpha \in [0, 1]$ and then $\beta = (1 - \alpha) / 2$. \label{pg:rev2_p7}This, in turn, simplifies the process of specifying a prior.

It is easy to verify that the stationary distribution $\vect{\pi}$ associated with this rate matrix is given by
\begin{equation}\label{eq:pi_ry56b}
\pi_i = \dfrac{- \alpha^2 + (5 - \alpha) \rho_i + (3 \alpha - 1) \rho_j - \alpha + 2}{2(3 - 2 \alpha) (\alpha+2)}, \qquad j = i + (-1)^{i+1}
\end{equation}
for $i=1,\ldots,4$. For ease of interpretation, it might seem more natural to reparameterise the model directly in terms of $\alpha \in [0,1]$ and the stationary distribution $\vect{\pi} \in \mathscr{S}_4$. However, given a fixed value for $\alpha$, the mapping \mbox{$\pi_{\alpha} : \mathscr{S}_4 \rightarrow \mathscr{S}_4$, where $\pi_{\alpha} (\vect{\rho}) = \vect{\pi}$,} is not surjective. This would substantially complicate inference. We therefore retain the original parameterisation, in terms of $\alpha$ and $\vect{\rho}$. 

Seeking an interpretation of $\vect{\rho}$, the relationship between each $\pi_i$ and the corresponding $\rho_i$ is complicated by the simplex constraints, which preclude isolation of the effect of a change in $\rho_i$ on $\pi_i$, whilst all the other $\rho_j$ remain fixed. However, from~\eqref{eq:pi_ry56b}, because $5 - \alpha > 3 \alpha - 1$, it is clear that for any fixed $\alpha \in [0,1]$, there is a positive linear relationship between, say, $\rho_1$ and $\pi_1$. The slope and intercept depend on how a simplex-preserving decrease in $\rho_2 + \rho_3 + \rho_4$ is shared between $\rho_2$ and $\rho_3 + \rho_4$ when $\rho_1$ is increased. To illustrate the relationships numerically, we simulate a sample of $\vect{\rho}$ vectors from a uniform distribution over $\mathscr{S}_4$ and then compute the corresponding stationary distribution $\vect{\pi}$ for various values of $\alpha$. Plots of $\pi_i$ against $\rho_i$ are displayed in Figure~S1 of the Supplementary Materials and show a strong positive relationship. We therefore interpret the parameter vector $\vect{\rho}$ as playing a role similar to the stationary distribution $\vect{\pi}$ in the HKY85 model. The parameter $\alpha$ then allows for differences between the rates of transition and transversion.

\subsection{\label{subsec:ry88a}The RY8.8a model}
The structure of the RY5.6b model is biologically appealing because of its simplicity and parallels with the widely used HKY85 model. However, it suffers a number of drawbacks. First, the model only has five degrees of freedom, which makes it inflexible compared with more complex Lie Markov models. Second, the additive structure of the instantaneous rates of change in the RY5.6b, as well as various other Lie Markov models, can often cause problems in the analysis of biological data. In many alignments, the empirical proportions of \texttt{A}, \texttt{G}, \texttt{C} and \texttt{T} are all reasonably close to 0.25 \citep[][]{BEPBS17}. If we imagine that the data arose from a stationary CTMP, this would demand $\pi_i \simeq 1/4$ for all $i=1,\ldots,4$. With reference to the RY5.6b model, with stationary distribution~\eqref{eq:pi_ry56b}, arguments of symmetry imply that for any $\alpha \in [0,1]$, we can only achieve $\pi_i = 1/4$ for all $i$ if $\rho_i = 1/4$ for all $i$. In this case, the ratio of the rates of change for transitions and transversions is given by
\begin{equation*}
(\alpha + 1/4) / \{ (1 - \alpha) / 2 + 1/4 \} = (4 \alpha + 1) / (3 - 2 \alpha) \le 5.
\end{equation*}
\begin{sloppypar}
\noindent However, experience suggests that for some alignments, we would expect a value much larger than this \citep[][]{RSK03}. In mammalian genomes, for example, this can occur due to 5-methylcytosine deamination to thymine at some sites, causing high rates of \texttt{C} to \texttt{T} point mutation \citep[][]{HE11}. This provides a possible explanation for the conclusions drawn in \citet{woodhams}, based on analyses of a large number of data sets, that the fit of the RY5.6b model is notably worse than that of the structurally similar HKY85 model. We therefore investigate a second (non-reversible) Lie Markov model, the RY8.8a model, which is more highly parameterised than RY5.6b and free from its additive structure. It has also been found to fit well in analyses of biological data \citep[][]{woodhams}.
\end{sloppypar}

As its name suggests, the RY8.8a rate matrix is based on the symmetry condition of purine-pyrimidine pairing and has eight degrees of freedom which can be represented by eight non-negative parameters. A representation of its rate matrix is given in Figure~\ref{fig:Q3}
where $\rho_1, \rho_2, \rho_3, \rho_4, \tilde{\rho}_5, \tilde{\rho}_6, \tilde{\rho}_7, \tilde{\rho}_8 \geq 0$. In order to fix the scale of the rate matrix, it is convenient to fix the trace as -1, then we can take $\rho_i = 2 \tilde{\rho}_i$ for $i=5,\ldots,8$, and restrict $\vect{\rho} \in \mathscr{S}_8$.

The analytic forms for the stationary probabilities $\vect{\pi}$ are given in Section~S1.2 of the Supplementary Materials. Like for the RY5.6b model, direct parameterisation in terms of $\vect{\pi}$ and, say, the rates of transition $(\rho_1, \rho_2, \rho_3, \rho_4)$, or the (scaled) rates of transversion $(\rho_5, \rho_6, \rho_7, \rho_8)$, would complicate inference because for fixed $(\rho_{i+1}, \rho_{i+2}, \rho_{i+3}, \rho_{i+4})$ where $i=0$ or $i=4$, we cannot invert the mapping from the remaining elements in $\vect{\rho} \in \mathscr{S}_8$ to $\vect{\pi} \in \mathscr{S}_4$. However, the parameters in the RY8.8a model have clear interpretations as instantaneous rates of change between different pairs of nucleotides and so we parameterise the model in terms of the single, interpretable stochastic vector $\vect{\rho}$.

\subsection{\label{subsec:nonhomo}Non-homogeneous RY5.6b and RY8.8a models}

As explained in Section~\ref{s:intro}, there are often both theoretical and empirical arguments for building non-stationarity into models for substitutions in molecular sequences. We therefore propose non-homogeneous and non-stationary extensions of the (non-reversible) RY5.6b and RY8.8a models outlined in the previous sections. A bifurcating rooted tree on $n$ taxa has $B = 2n-2$ branches; the underpinning unrooted topology has one fewer. Following the framework outlined in Section~\ref{s:intro}, we construct a non-homogeneous RY5.6b model by allowing evolution along every branch $b$ of the associated unrooted topology to be controlled by its own rate matrix $\matr{Q}_b$ which belongs to the RY5.6b family. Rooting this tree on branch $r$ of the unrooted topology divides the branch into two. The rate matrix $\matr{Q}_r$ is associated with the two new branches on either side of the root, whilst its stationary distribution is used as the distribution at the root. We define our non-homogeneous RY8.8a model in an analogous fashion. \label{pg:rev1_p7}It is worth mentioning that an alternative, though less parsimonious, way to formulate the models would be to allow the branches on either side of the root to have their own rate matrix with a simplex-valued parameter describing the distribution at the root of the tree.  

Computational inference is greatly simplified if the number of parameters which vary from branch to branch is kept small. In earlier work, where we developed non-homogeneous, non-stationary extensions of the (reversible) HKY85 and GTR models \citep[][]{heaps,williams15}, this was achieved by keeping the exchangeability parameters fixed across the tree, so that only the stationary probabilities varied. For the RY5.6b model, we take a similar approach, allowing only the parameter $\vect{\rho} \in \mathscr{S}_4$, which serves as a proxy for the stationary distribution, to vary across branches. The parameter $\alpha$, controlling the differences  between the rates of transition and transversion, is held constant. For the RY8.8a model, there is no corresponding partition of the parameters, and so we allow all parameters in $\vect{\rho} \in \mathscr{S}_8$ to vary from branch to branch.

This yields non-homogeneous RY5.6b and RY8.8a models in which a set of branch-specific simplex-valued parameters $\{ \vect{\rho}_1, \ldots, \vect{\rho}_{B-1} \}$ induce corresponding heterogeneity in the theoretical stationary distribution across branches. The models are therefore non-stationary, with step-changes in the stationary distribution at each speciation event. This allows us to capture change in sequence composition over evolutionary time.

Our non-homogeneous, non-stationary, locally non-reversible models offer two main advantages over their locally reversible counterparts. First, as we investigate further in Section~\ref{sec:simulations}, the property of non-reversibility can provide an additional source of likelihood information about the direction of time, and hence the position of the root. Second, if we prune $n_0$ taxa from a tree on $n$-species, the non-homogeneous Lie Markov model on DNA characters in $\Omega^n$ induces a distribution on the reduced DNA characters in $\Omega^{n-n_0}$. Because Lie Markov models are closed under matrix multiplication, this distribution could, in \emph{most} cases, have been constructed directly from a non-homogeneous Lie Markov model over the $n-n_0$-taxa subtree. (We note that this cannot be guaranteed in \emph{all} cases because it is theoretically possible for the product of two Lie Markov rate matrices $Q_1,Q_2 \in \mathcal{M}$, to yield a rate matrix $Q = \log \{ \exp(Q_1) \exp(Q_2) \} \in \mathcal{M}$ which is not stochastic; see~\citet{woodhams} for an empirical investigation.)\label{pg:rev1_p8} Non-homogeneous, non-stationary but locally reversible models lack this property of mathematical consistency.

\section{Prior distribution for substitution model parameters}
\label{sec:prior}
In a homogeneous model, the instantaneous rate matrix which characterises the evolutionary process is the same on all branches of the phylogeny. In our non-homogeneous models, it can change from branch to branch. Letting $K=4$ and $K=8$ for the RY5.6b and RY8.8a models, respectively, we adopt a prior in which the branch-specific parameter vectors, $\vect{\rho}_1, \ldots, \vect{\rho}_{B-1} \in \mathscr{S}_K$, are positively correlated. This provides flexibility, whilst retaining some of the benefits of the homogeneous model, by allowing information to be shared between branches. As we move from one branch to its descendants, we do not anticipate a substantial change in the evolutionary process. We therefore build explicit dependence on recent ancestors into our joint prior through the assignment of a stationary, first order autoregression over a reparameterised set of vectors $\vect{\varrho}_b \in \mathbb{R}^{K-1}$, $b=1,\ldots,B-1$, each of which is related to the corresponding $\vect{\rho}_b \in \mathscr{S}_K$ through a linear mapping, followed by multinomial logit transformation. Full details of the reparameterisation and prior are provided in \citet{heaps} but, briefly, its role is to induce a distribution for the $\vect{\rho}_b$ which is symmetric with respect to its $K$ components, and has common marginal mean and variance for all $b=1,\ldots,B-1$. By construction, the prior for the reparameterised vectors depends on the tree topology $\tau$ and is given by
\begin{equation*}
\pi(\vect{\varrho}_1,\dots,\vect{\varrho}_{B-1} | \tau) = \textstyle\prod_{k=1}^{K-1} \left\{ \pi(\varrho_{rk} | \tau) \prod_{b \ne r} \pi(\varrho_{bk} | \varrho_{a(b),k}, \tau) \right\},
\end{equation*}
where $r$ is the index of the rooting branch and $a(b)$ is the index of the ancestral branch (or root) of branch $b$. Then for $k=1,\ldots,K-1$ we have $\varrho_{rk} | \tau \sim \norm\left(0\,,\,v_{\varrho}/(1-p_{\varrho}^2)\right)$ and, for $b \ne r$, $\varrho_{bk} | \varrho_{a(b),k}, \tau \sim \norm(p_{\varrho} \varrho_{a(b),k}, v_{\varrho})$ in which $p_{\varrho} \in [0,1]$ and $v_{\varrho} \in \mathbb{R}^+$ are fixed hyperparameters that control the marginal variances and correlations of the $\vect{\varrho}_b$ and hence $\vect{\rho}_b$.

In the RY5.6b model, the instantaneous rate matrix $\matr{Q}_b$ on branch $b$ depends on the parameter $\alpha \in [0,1]$ in addition to the stochastic vector $\vect{\rho}_b \in \mathscr{S}_4$. Conditional on $\tau$, we factorise the joint prior of $\alpha$ and the $\vect{\rho}_b$ as $\pi(\alpha, \vect{\rho}_1, \ldots, \vect{\rho}_{B-1} | \tau) = \pi(\alpha) \pi(\vect{\rho}_1, \ldots, \vect{\rho}_{B-1} | \tau)$ and assign a flat distribution to $\alpha$, that is, $\alpha \sim \bet(1, 1)$.

\section{\label{sec:posterior}Posterior inference via MCMC}

The unknowns in the model comprise the rooted tree topology $\tau$, branch lengths $\vect{\ell} = (\ell_1, \ldots, \ell_B)^T \in \mathbb{R}^B_+$ and the shape parameter $\phi \in \mathbb{R}_+$ in the discretised gamma distribution for rate variation across sites. We also have the set of substitution model parameters, which we denote by $\mathcal{Q}$, where $\mathcal{Q} = \{ \alpha, \vect{\varrho}_1, \ldots, \vect{\varrho}_{B-1} \}$ for the non-homogeneous RY5.6b model and $\mathcal{Q} = \{ \vect{\varrho}_1, \ldots, \vect{\varrho}_{B-1} \}$ for the non-homogeneous RY8.8a model. These parameters determine the distribution at the root of the tree, say $\vect{\pi}_0$, and the instantaneous rate matrices, $\matr{Q}_1, \ldots, \matr{Q}_B$, on each branch.  

The posterior distribution for the unknowns can be expressed as\\ \mbox{$\pi(\tau, \vect{\ell}, \phi, \mathcal{Q} | \matr{y}) \propto p(\matr{y}| \tau, \vect{\ell}, \phi, \mathcal{Q}) \pi(\tau, \vect{\ell}, \phi, \mathcal{Q})$}, in which $p(\matr{y}| \tau, \vect{\ell}, \phi, \mathcal{Q})$ is the likelihood of the alignment $\matr{y}$ and $\pi(\tau, \vect{\ell}, \phi, \mathcal{Q})$ is the prior density. The likelihood is calculated as $p(\matr{y}| \tau, \vect{\ell}, \phi, \mathcal{Q}) = \prod_{i=1}^m p(\vect{Y}_i = \vect{y}_i| \tau, \vect{\ell}, \phi, \mathcal{Q})$ in which $\vect{Y}_i \in \Omega^n$ is the DNA character at site $i$. The probability of the observed character $\vect{y}_i$ at site $i$ is given by  
\begin{equation*}
\Pr(\vect{Y}_i = \vect{y}_i | \tau, \vect{\ell}, \phi, \mathcal{Q}) = \frac{1}{4} \sum_{k=1}^4 \sum_{X} \pi_{0,X(0)} \prod_{\text{edges} \hspace*{3.0pt} b = (v,w)} p_{b, X(v), X(w)} \{ r_k(\phi) \ell_b \}.
\end{equation*}
Here $v$ and $w$ are the vertices (nodes) at the two ends of edge $b$ with length $\ell_b$, $X(u)$ is the character at vertex $u$, $u=0$ denotes the root vertex and $P_b\{ r_k(\phi) \ell_b \} = [ p_{bhi}\{ r_k(\phi) \ell_b \} ] = \exp \{ r_k(\phi) \ell_b \matr{Q}_b' \}$ is the transition matrix associated with edge $b$ for discretised site rate category $k$. The inner sum is over all functions $X$ from the vertices to $\Omega$ such that $X(u)$ matches the data $y_i(u)$ for all leaf vertices $u$. It can be computed efficiently using a post-order traversal of the tree called Felsenstein's pruning algorithm \citep[][]{felsenstein}. The outer sum is over the four rate categories of the discretised gamma distribution for rate variation across sites.

The posterior density $\pi(\tau, \vect{\ell}, \phi, \mathcal{Q} | \matr{y})$ is not available analytically. We therefore build up a numerical approximation by generating samples from the posterior using a Metropolis within Gibbs sampling scheme which iterates through a series of updates for each unknown. Parameters which lie in $\mathbb{R}$ or $\mathbb{R}_+$, can be updated using standard proposal distributions, for example Gaussian random walks for the reparameterised branch-specific parameters $\vect{\varrho}_b$. For the parameter $\alpha \in [0,1]$ in the non-homogeneous RY5.6b model we generate proposals $\alpha^{\ast}$ from a Beta distribution which is roughly centred at the current value $\alpha$, namely \mbox{$\alpha^{\ast} | \alpha \sim \bet \left( s_1 \alpha + s_2 \, ,  \, s_1(1 - \alpha) + s_2 \right)$}. Here $s_1 \in \mathbb{R}_+$ and $s_2 \in \mathbb{R}_+$ are tuning parameters. The first affects the variance of the proposal and should be tuned to adjust the acceptance rate. The second helps to prevent the sampler from sticking at the boundaries of the unit interval and should be set close to zero; for example, $s_2=0.005$.

Finally, the rooted topology $\tau$ can be updated using standard proposals for topological moves such as nearest neighbour interchange (NNI), subtree prune and regraft (SPR), and proposals to alter the root position; see \citet{heaps} for a complete description of all three moves. The variation across branches in the parameter vectors $\vect{\varrho}_b$ complicates these topological proposals because they must include modifications to the $\vect{\varrho}_b$, as well as branch lengths $\ell_b$, for the edges whose local interpretation is changed by the proposal. To generate such proposals, we can, for example, propose the new $\vect{\varrho}_b$ using a distribution centred at the parameter vector on a neighbouring branch. The MCMC inferential procedures are programmed in Java. A software implementation can be found in the Supplementary Material.

\section{\label{sec:simulations}Analysis of simulated data}

\label{pg:ed_1}For the results of model-based inference on the root position to offer biological insight, the position of the root has to be identifiable under the likelihood. Proving that this is the case for models that are non-stationary, non-reversible, or both, is extremely challenging, except in very special cases \citep[][]{kaehler2017}. On the other hand, carefully designed simulation experiments can readily be used to provide empirical evidence of identifiability, and to investigate the conditions under which inference more closely reflects the data-generating mechanism. We therefore adopt a simulation-based approach to investigate the identifiability of the root position and underlying topology in our non-homogeneous RY5.6b and RY8.8a models. Specifically, we consider the effects of: (i) different numbers of taxa and sites; (ii) different topologies and branch lengths.

In all simulations, our prior for the non-homogeneous RY5.6b model takes the form 
\begin{equation*}
\pi(\tau, \vect{\ell}, \phi, \alpha, \vect{\varrho}_1, \ldots, \vect{\varrho}_{B-1}) = \pi(\tau) \pi(\phi) \pi(\alpha, \vect{\varrho}_1, \ldots, \vect{\varrho}_{B-1} | \tau) \prod_{b=1}^B \pi(\ell_b)
\end{equation*}
in which $\tau$ denotes the rooted topology and $\vect{\ell} = (\ell_1, \ldots, \ell_B)^T \in \mathbb{R}^B_+$ denotes the branch lengths. The prior for the non-homogeneous RY8.8a model has the same structure, but omitting the parameter $\alpha$. We assign priors $\ell_b \sim \expo(10)$ to the branch lengths and a distribution $\phi \sim \gam(10, 10)$ to the shape parameter in the gamma distribution for rate heterogeneity across sites. The rooted topology is given a prior according to the Yule model of speciation. Defining a root split of size $j  \! :  \! (n-j)$, $j \in \{ 1, \ldots, \lfloor n/2 \rfloor \}$, as the set of all rooted trees with $j$ taxa on one side of the root and $n-j$ on the other, the Yule model generates a distribution in which near equal probability is assigned to root splits of all sizes \citep[][]{cherlin17}. Finally, the priors $\pi(\alpha, \vect{\varrho}_1, \ldots, \vect{\varrho}_{B-1} | \tau)$ for RY5.6b and $\pi(\vect{\varrho}_1, \ldots, \vect{\varrho}_{B-1} | \tau)$ for RY8.8a were described in detail in Section~\ref{sec:prior}. The choices of the hyperparameters $p_{\varrho}$ and $v_{\varrho}$ in these priors are given and justified in Section~S3.1 of the Supplementary Materials.

For each analysis, we used the MCMC algorithm described in Section~\ref{sec:posterior}. Two chains, initialised at different starting points, were each allowed to run for 1M iterations, discarding the first 500K as burn-in and thinning the remaining output to retain every 100-th iteration so as to reduce computational overheads. The standard graphical and numerical diagnostics used in phylogenetic inference \citep[][]{PhyloBayesManual} were used to assess convergence and mixing.

\label{pg:rev1_p10}Carrying out computation on one server with two six-core Xeon E5645 CPUs and 16GB RAM, the time required to generate 1M iterations from an alignment with 12 taxa and 1000 sites was around 3.5 days for both models. Section~S3.2 of the Supplementary Materials contains further discussion on how this computational time compares with that for simpler substitution models and how it scales with the number of sites and taxa.

\subsection{\label{subsec:taxa_and_sites}Different numbers of taxa and sites}

In order to assess the extent to which root inference depends upon the dimensions of the data being analysed, we simulated alignments under the non-homogeneous RY5.6b and RY8.8a models, varying the number of taxa ($6$, $12$, $24$) and the number of sites ($500$, $1000$, $2000$). First an unrooted tree on 24 taxa was simulated by random resolution of a star tree. This was rooted to form a balanced tree, that is, a tree with an equal number of taxa on either side of the root, and then branch lengths were sampled from a $\gam(2,20)$ distribution. The branch-specific parameter vectors $\vect{\rho}_b$ were simulated from Dirichlet $\dir_4(10, 10, 10, 10)$ and Dirichlet $\dir_8(3,3,3,3,3,3,3,3)$ distributions for the RY5.6b and RY8.8a models, respectively. This gave a degree of heterogeneity which was consistent with what we have seen in our analyses of biological data. For the non-homogeneous RY5.6b model, $\alpha$ was set at $0.5$, the mean of its symmetric prior. Likewise, the shape parameter $\phi$ in the discretised gamma distribution for rate variation across sites was set to the mode of its prior, $0.9$. Using this rooted tree and these parameter values, three alignments of $2000$ sites were simulated under each model.  Taxa, sites, or taxa and sites were then removed from the alignments to give three data sets for each combination of sites and taxa specified above. The taxa that are pruned were chosen uniformly at random, but constrained so that the corresponding tree for each resulting alignment is balanced. This was to avoid any potential confounding with the effect of balance in the rooted topology, which we examine separately in Section~\ref{subsec:topologies_and_BLs}. The rooted topologies on $6$, $12$ and $24$ taxa are displayed in Supplementary Figure~S2. 

\begin{figure}[!p]
\centering
\subfloat[\label{fig:m5_6_root}]{\includegraphics[width=0.33\textwidth]{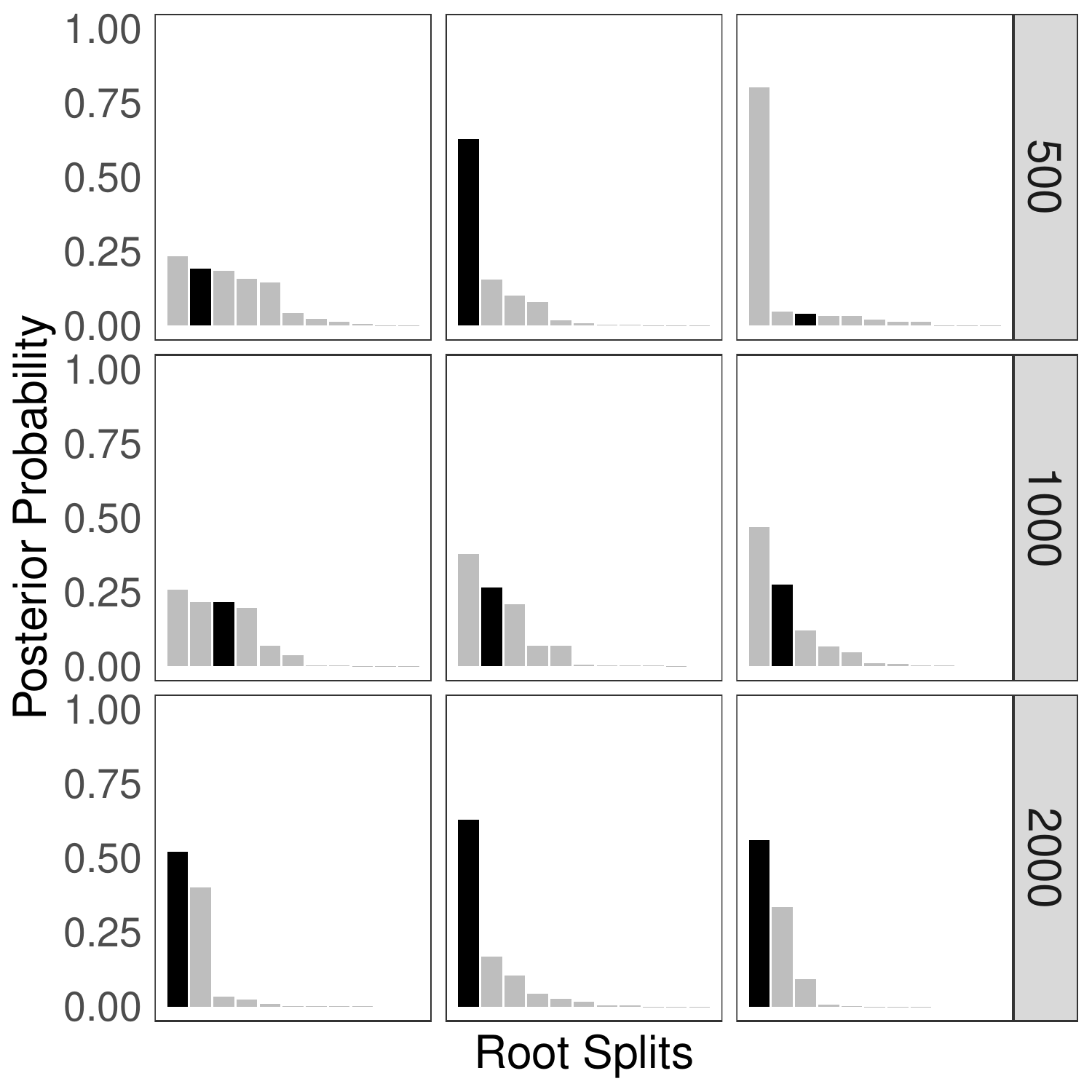}}
\hspace{2cm}
\subfloat[\label{fig:m6_6_root}]{\includegraphics[width=0.33\textwidth]{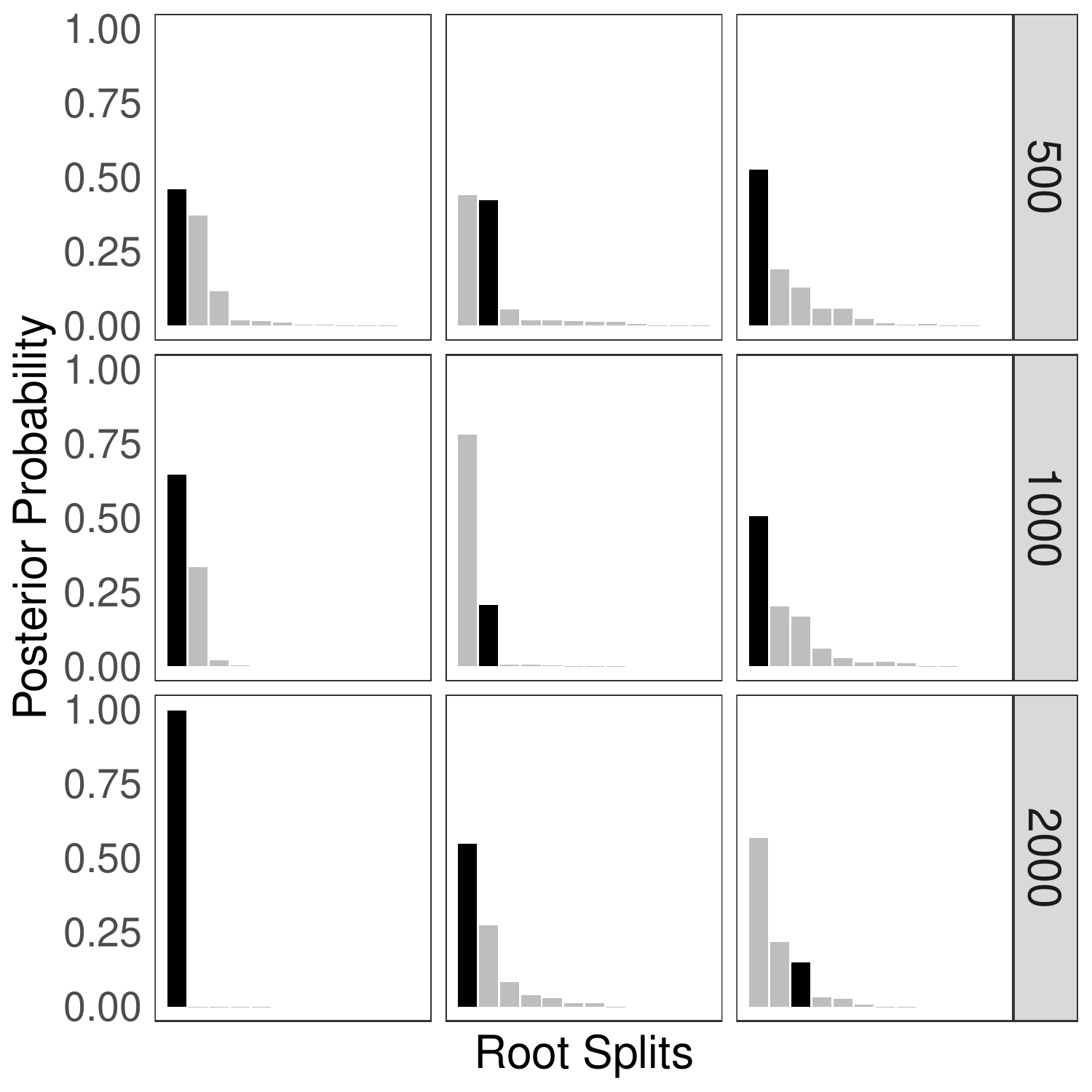}}\\
\subfloat[\label{fig:m5_12_root}]{\includegraphics[width=0.33\textwidth]{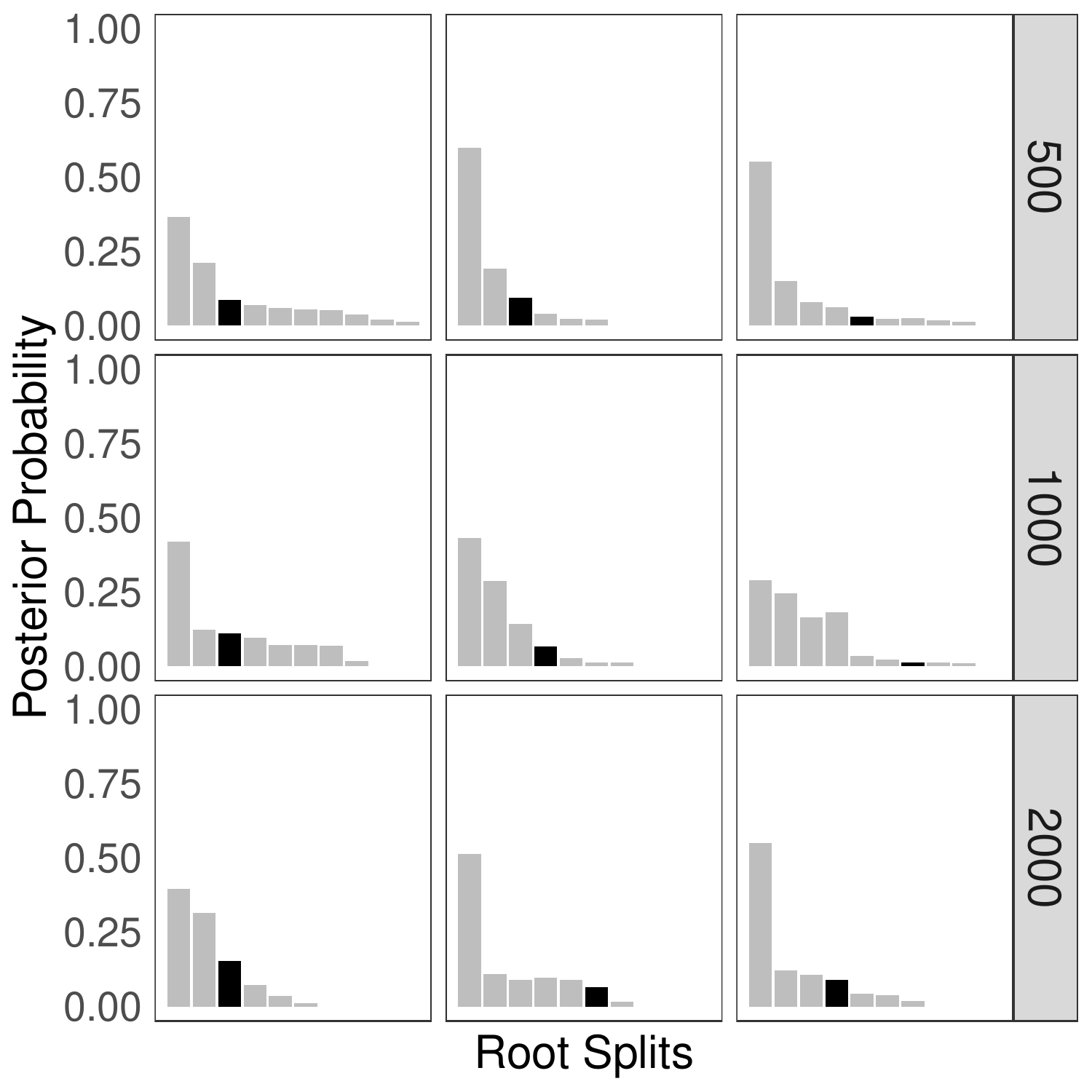}}
\hspace{2cm}
\subfloat[\label{fig:m6_12_root}]{\includegraphics[width=0.33\textwidth]{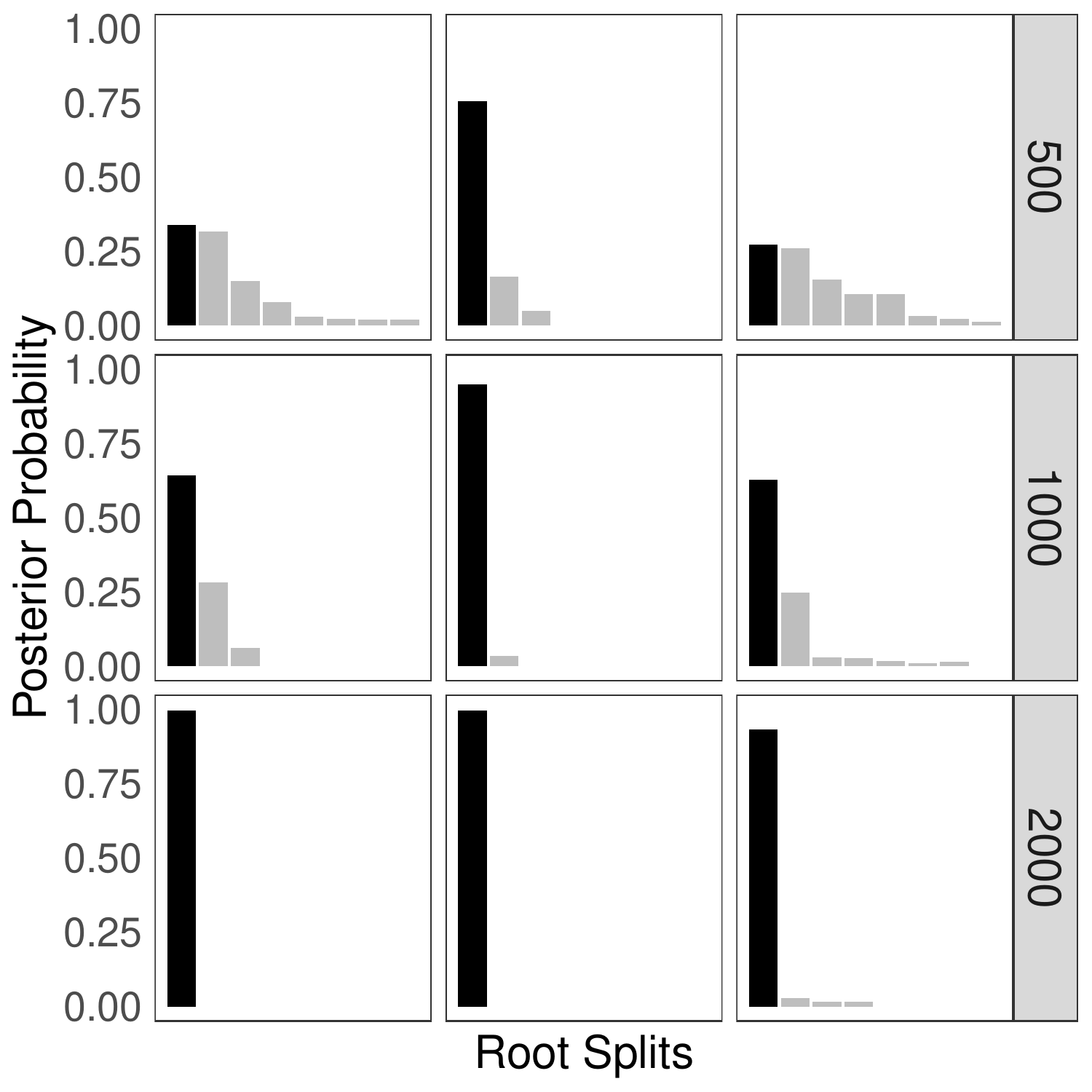}}\\
\subfloat[\label{fig:m5_24_root}]{\includegraphics[width=0.33\textwidth]{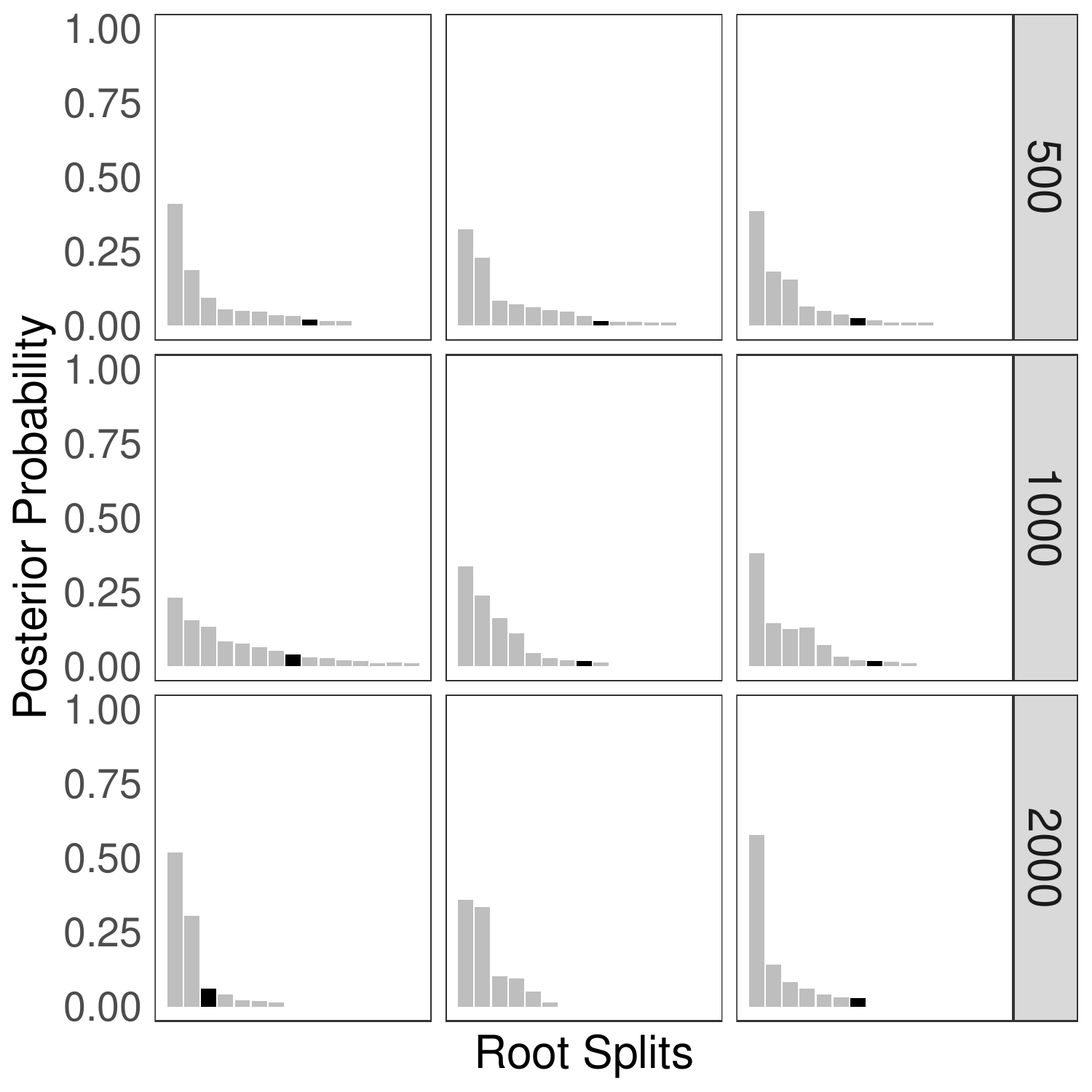}}
\hspace{2cm}
\subfloat[\label{fig:m6_24_root}]{\includegraphics[width=0.33\textwidth]{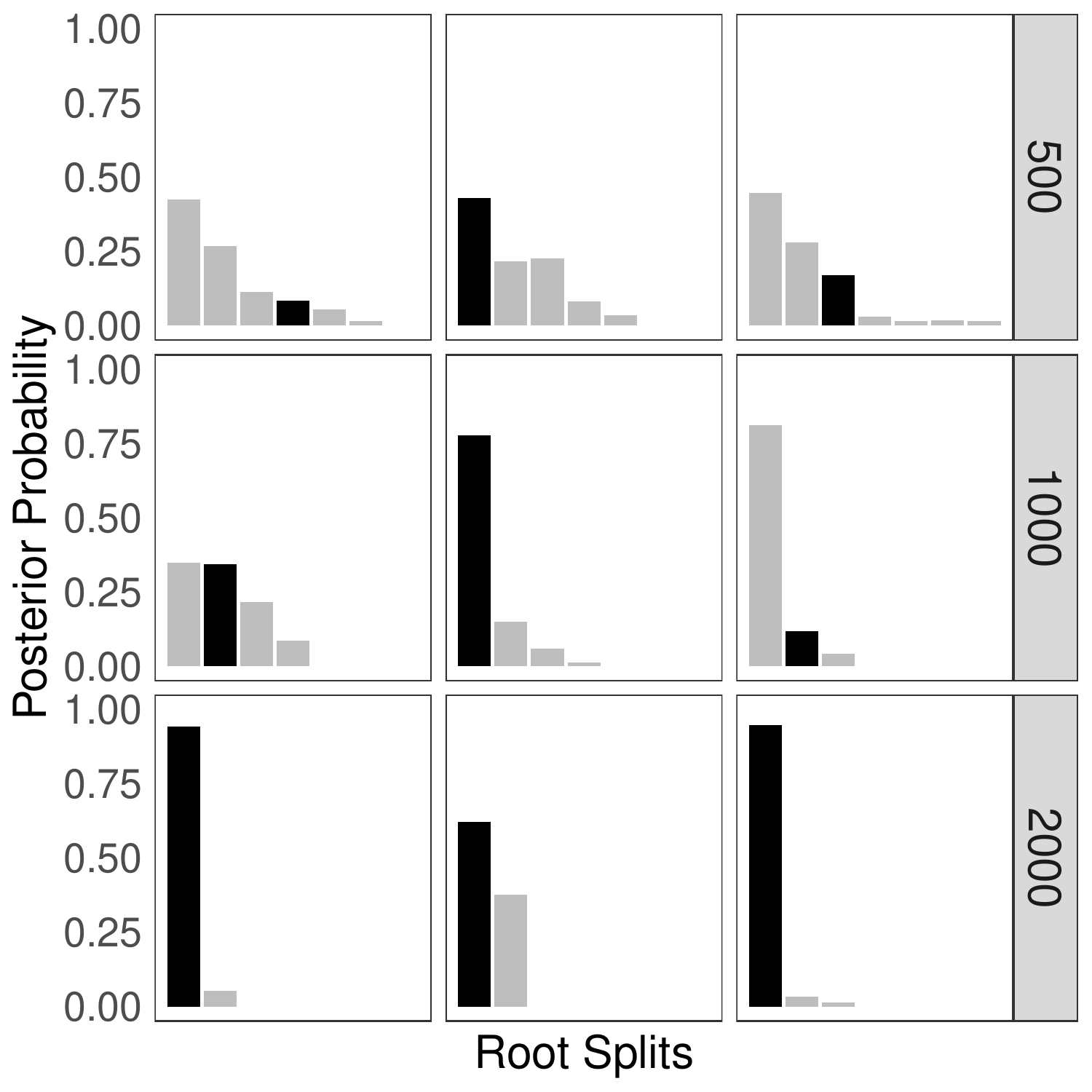}}
\caption{Posterior distribution over roots splits when three data sets are simulated and analysed under the non-homogeneous RY5.6b model and the number of taxa is \protect\subref{fig:m5_6_root} $6$, \protect\subref{fig:m5_12_root} $12$, \protect\subref{fig:m5_24_root} $24$; and when three data sets are simulated and analysed under the non-homogeneous RY8.8a model and the number of taxa is \protect\subref{fig:m6_6_root} $6$, \protect\subref{fig:m6_12_root} $12$, \protect\subref{fig:m6_24_root} $24$. In every plot, bars are arranged in descending order of posterior probability and the correct root split is highlighted in black. In the plots for 12 and 24 taxa, bars corresponding to probabilities less than $0.01$ have been removed to improve readability.}\label{fig:taxa_and_sites_root}
\end{figure}


For the trees on $6$, $12$ and $24$ taxa, Figures~\ref{fig:m5_6_root}, \ref{fig:m5_12_root} and \ref{fig:m5_24_root}, respectively, display the posterior distribution over root splits for the alignments simulated and analysed under the non-homogeneous RY5.6b model. Supplementary Figures~S3a, S3c and S3e show the analogous plots for the posterior distribution over unrooted topologies. The black bars highlight the (true) root split or unrooted topology from the tree used to simulate the data. The corresponding plots for the non-homogeneous RY8.8a model are shown in Figures~\ref{fig:m6_6_root}, \ref{fig:m6_12_root} and \ref{fig:m6_24_root} and Supplementary Figures~S3b, S3d and S3e, respectively. As expected, irrespective of the model or number of taxa, the posterior support for the correct root split tends to increase as the number of sites increases, and the correct root split is more frequently identified as the posterior mode. The same is true for unrooted topologies. Indeed, when there are 2000 sites in the alignment and the non-homogeneous RY8.8a model is used, the posterior probability for the correct root split is close to one for all tree sizes and the posterior probability of the correct unrooted topology is 0.758 on average across the three alignments. Although increasing the number of taxa leads to quadratic growth in the number of possible root splits and super-exponential growth in the number of possible unrooted topologies, it does not seem to have a detrimental effect on inferential performance for the non-homogeneous RY8.8a model. Unfortunately, the same is not true for the RY5.6b model, under which inference of the root position is generally worse, particularly for larger trees. For example, when the number of taxa is $12$ or $24$, the correct root split is not recovered as the posterior mode in any simulations. The better rooting performance of the RY8.8a model is likely explained by two factors. First, the model has more parameters that can vary across the tree and induce non-stationary behaviour. \label{pg:rev2_identif2}Second, conditional on the correct rooted topology, investigation into the identifiability of the branch-specific parameters, $\vect{\rho}_1, \ldots, \vect{\rho}_{B-1}$, revealed that the $\vect{\rho}_b$ are better identified in the non-homogeneous RY8.8a model than the RY5.6b model; see Supplementary Figures~S4 and S5. This may be because the additive structure of the RY5.6b rate matrix makes the likelihood more flat and hence less sensitive to changes in the $\vect{\rho}_b$.

\subsection{\label{subsec:topologies_and_BLs}Different topologies and branch lengths}

In earlier work investigating a homogeneous, stationary, non-reversible model \citep[][]{cherlin17}, we found root inference to be sensitive to some of the prior-data conflicts that occur commonly in the analysis of biological data. Typically these arise due to incongruent prior and likelihood information about branch lengths and the rooted topology. In our analyses we adopt the near ubiquitous prior for the set of branch lengths, which structures beliefs as independent $\expo(10)$ distributions. This prior places 99.9\% of its mass below 0.691 and so asserts a strong belief that branch lengths are reasonably short. As a consequence, given an unrooted topology that contains a long branch, the prior supports rooting on this branch in order to split it into two shorter edges. In our analyses we adopt a Yule prior over rooted topologies. As discussed previously, one of the compelling properties of the Yule distribution is that it assigns near equal probability  to root splits of all sizes. However, a combinatorial consequence of this property is that more support is assigned to balanced than unbalanced trees. In this section we analyse simulated data to explore posterior sensitivity to prior-data conflicts that arise because of long branches in the underlying unrooted tree or an unbalanced rooted topology.

\begin{figure}[th]
\centering
\includegraphics[width=0.8\textwidth]{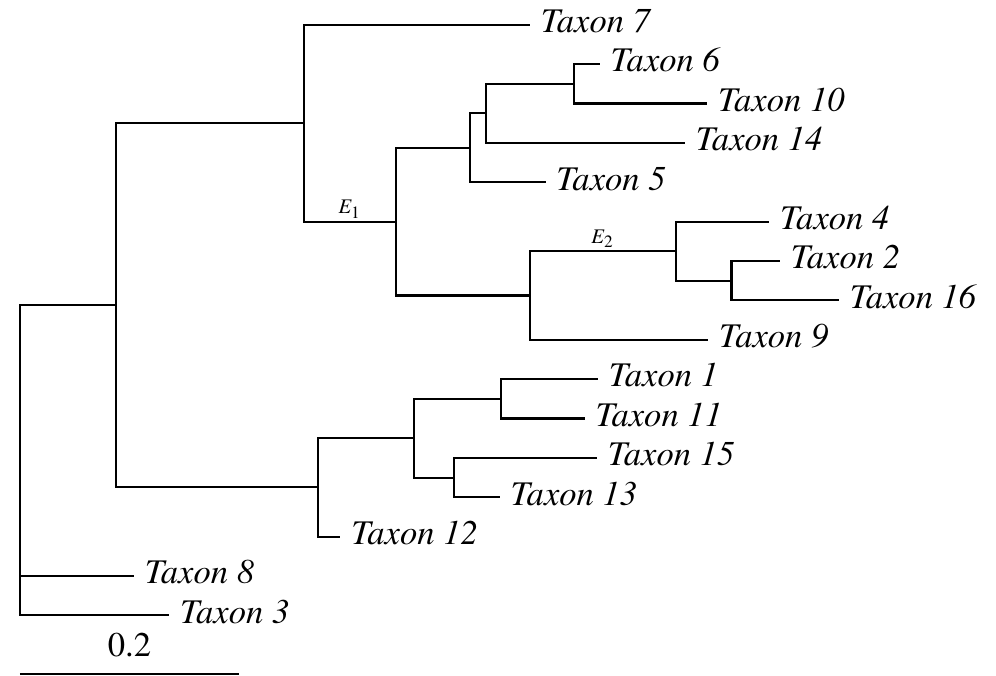}
\caption{Unrooted tree used in simulation experiments to investigate the effects of different topologies and branch lengths on root inference. In the experiments, the tree is rooted at the midpoint of either branch $E_1$ or $E_2$. The tree is depicted with branch $E_1$ having a ``medium'' length of $0.084$ units. In the experiment, this is varied to $0.237$ units (``long'') or $0.018$ units (``short'').}\label{fig:unrooted_tree}
\end{figure}

In order to facilitate straightforward comparison with our earlier work, we set up the simulation experiment in the same way as \citet{cherlin17}. Simulations were based on the unrooted tree on $16$ taxa depicted in Figure~\ref{fig:unrooted_tree} whose topology was simulated through random resolution of a star tree. Branch lengths were simulated from a $\gam(2, 20)$ distribution. Based on this unrooted tree we construct six different rooted trees by varying the root position, which is placed at the midpoint either of branch $E_1$ or branch $E_2$, and the length of the branch $E_1$, which is either the 95\%, 50\% or 5\% quantile of the $\gam(2, 20)$ distribution:
\begin{description}
\item[Tree $1$:] balanced (rooted on $E_1$), long root branch (length $0.237$);
\item[Tree $2$:] unbalanced (rooted on $E_2$), long internal branch (length $0.237$);
\item[Tree $3$:] balanced (rooted on $E_1$), short root branch (length $0.018$);
\item[Tree $4$:] unbalanced (rooted on $E_2$), short internal branch (length $0.018$);
\item[Tree $5$:] balanced (rooted on $E_1$), medium root branch (length $0.084$);
\item[Tree $6$:] unbalanced (rooted on $E_2$), medium internal branch (length $0.084$).
\end{description}
As indicated above, Trees $1$, $3$ and $5$ have a balanced rooted topology, with root type $8:8$, whilst Trees $2$, $4$ and $6$ are unbalanced, with root type $3:13$. The Yule prior offers more than six times more mass to the balanced tree and hence the prior and likelihood are likely to be in conflict when the tree is unbalanced. In the unrooted tree associated with Trees $1$ and $2$, branch $E_1$ is the longest, whilst for Trees $3$ and $4$ it is among the shortest. Given the unrooted topology depicted in Figure~\ref{fig:unrooted_tree}, the prior support for a root on edge $E_1$ increases as the branch becomes longer, and hence will increasingly conflict with the likelihood if $E_1$ is not the root edge.

\begin{figure}[!t]
\centering
\subfloat[\label{fig:m6_bl_root}]{\includegraphics[width=0.4\textwidth]{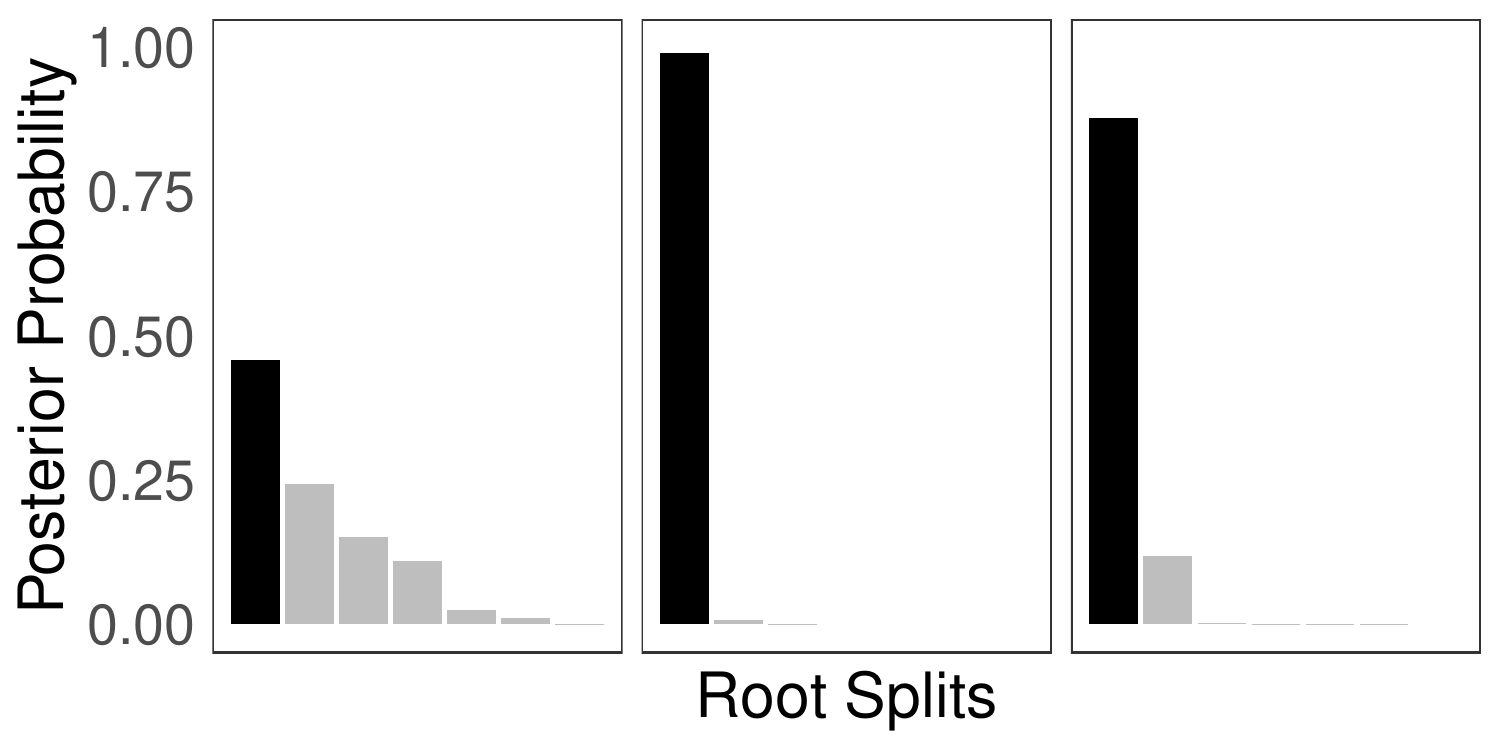}}
\hspace{1.5cm}
\subfloat[\label{fig:m6_ul_root}]{\includegraphics[width=0.4\textwidth]{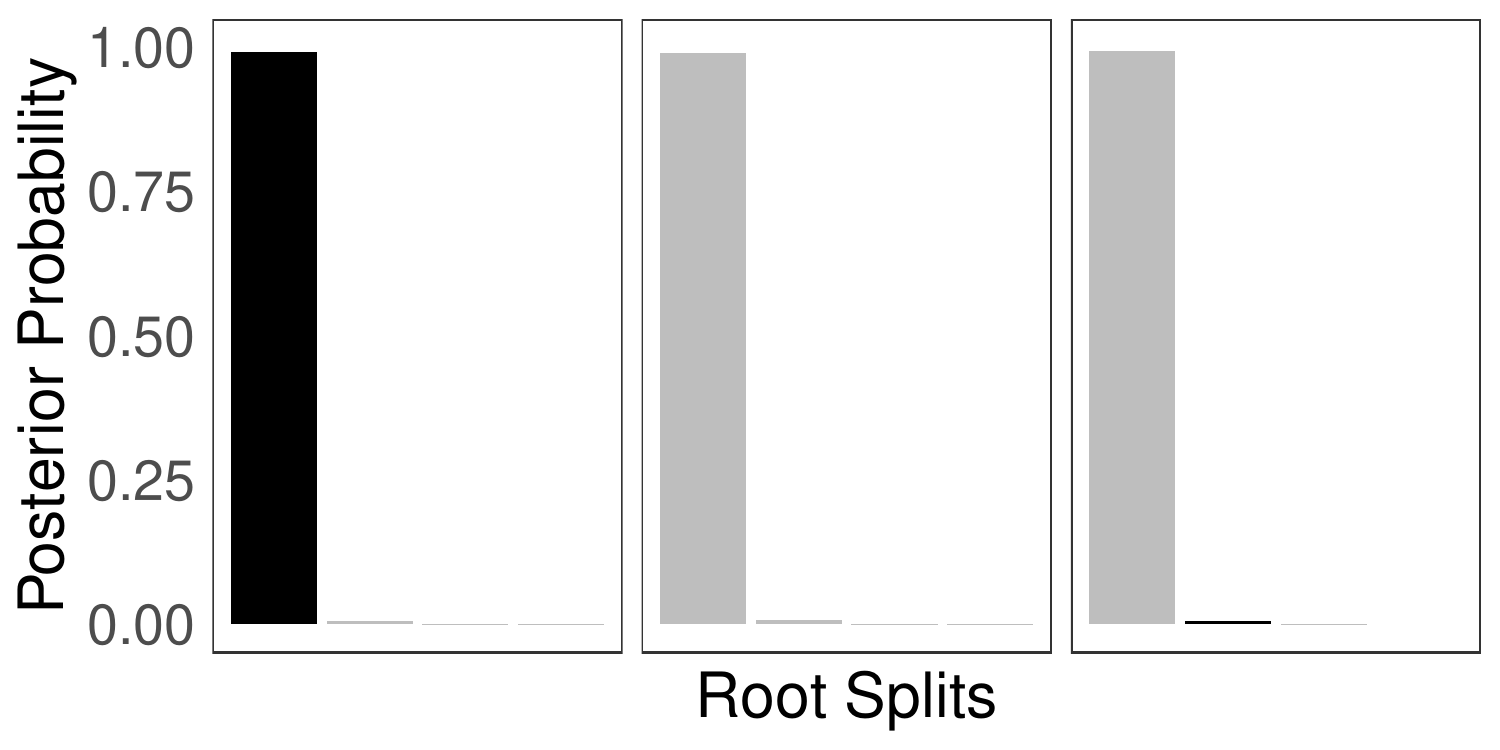}}\\
\subfloat[\label{fig:m6_bs_root}]{\includegraphics[width=0.4\textwidth]{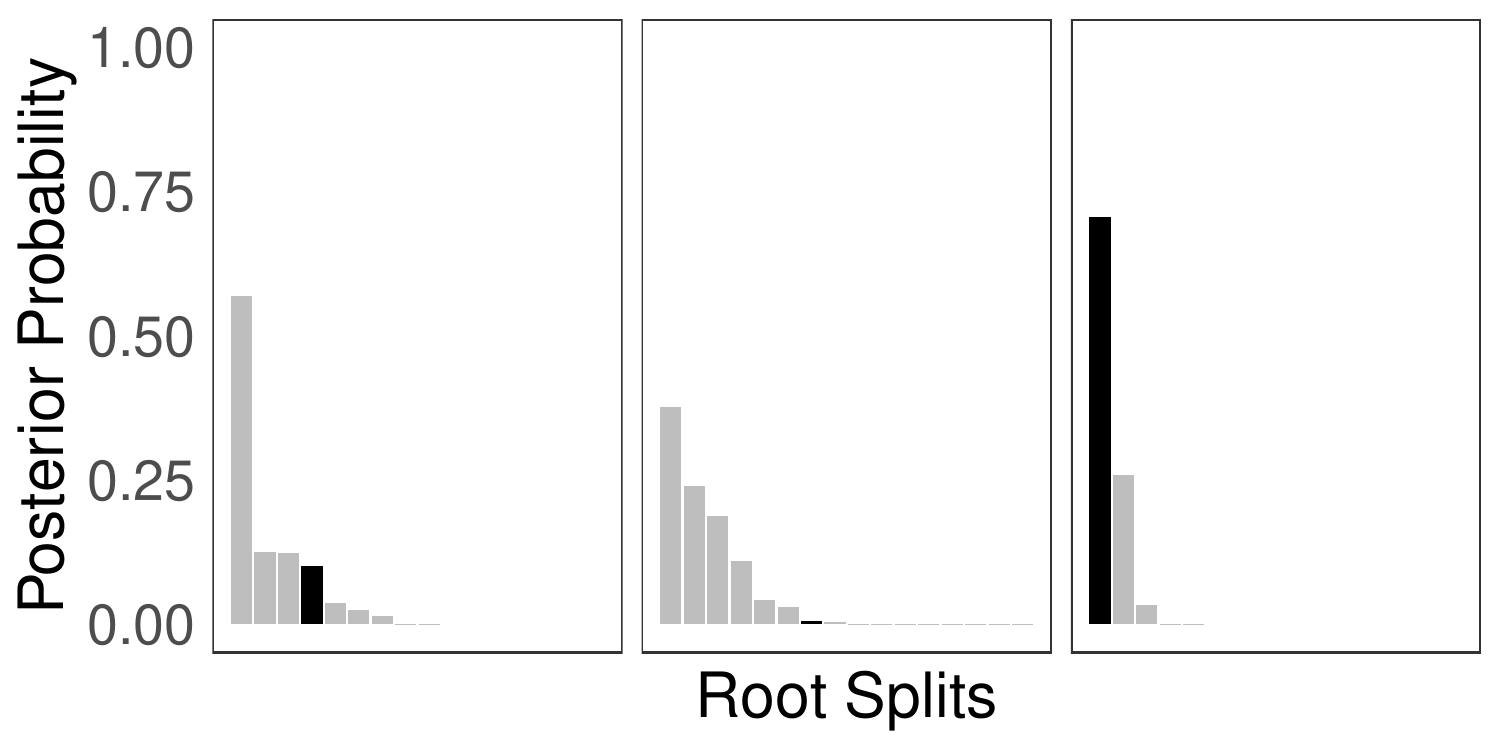}}
\hspace{1.5cm}
\subfloat[\label{fig:m6_us_root}]{\includegraphics[width=0.4\textwidth]{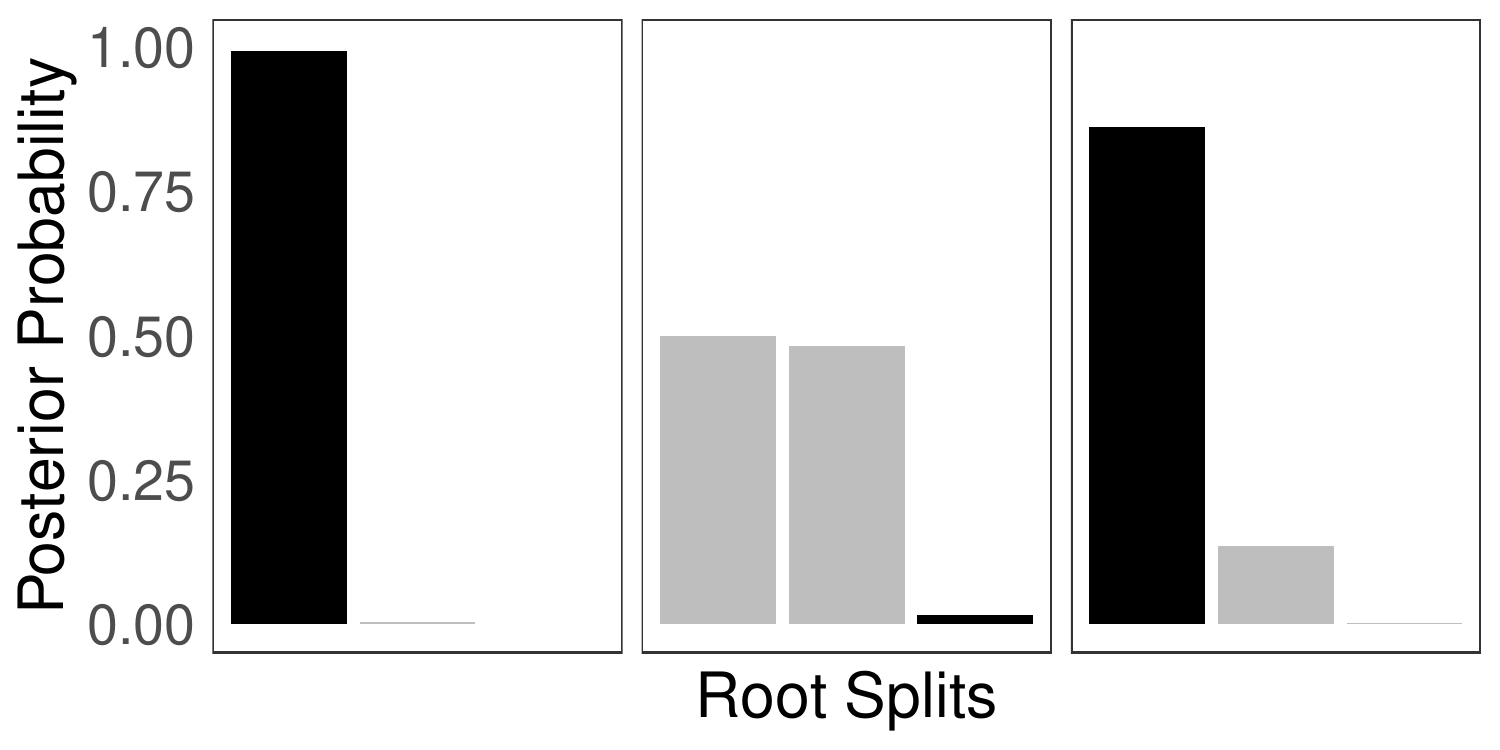}}\\
\subfloat[\label{fig:m6_bm_root}]{\includegraphics[width=0.4\textwidth]{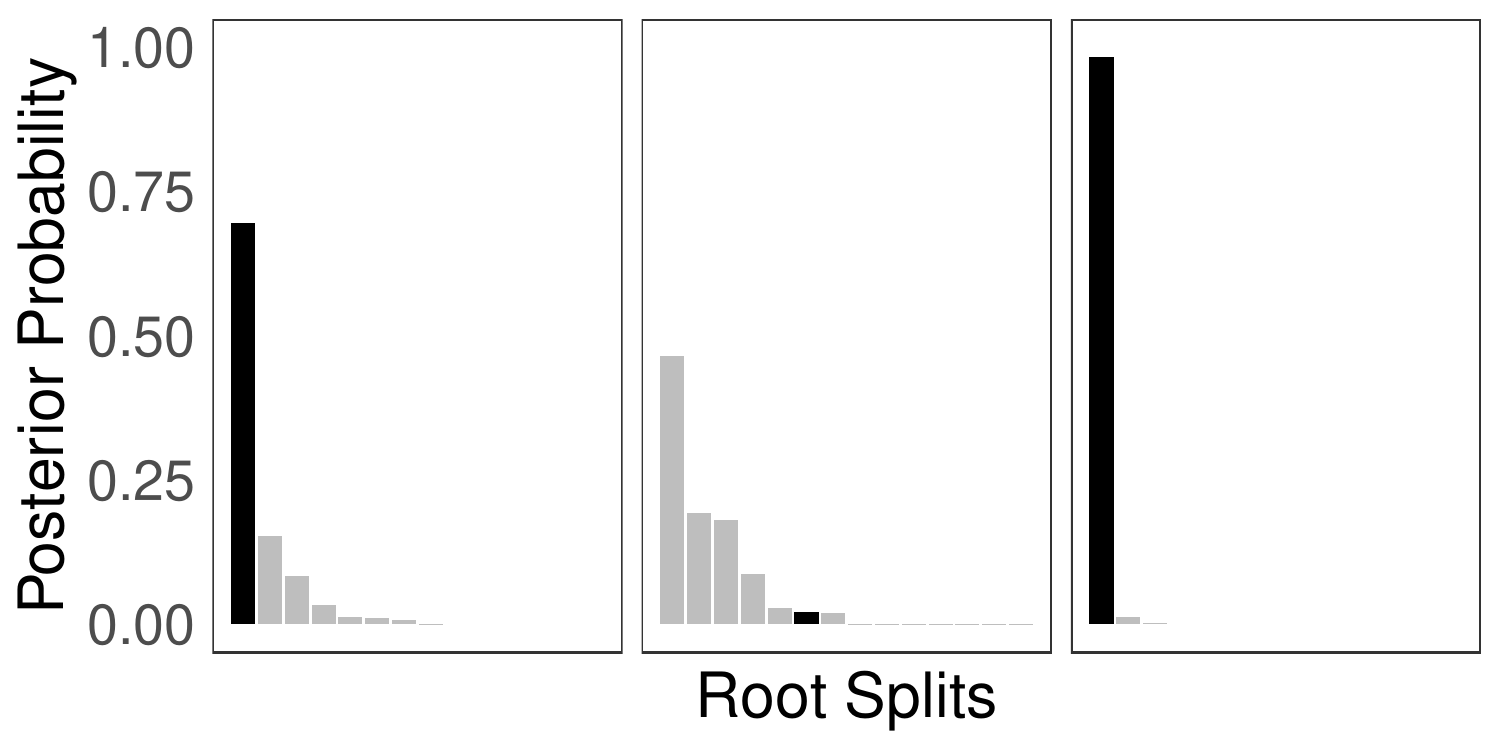}}
\hspace{1.5cm}
\subfloat[\label{fig:m6_um_root}]{\includegraphics[width=0.4\textwidth]{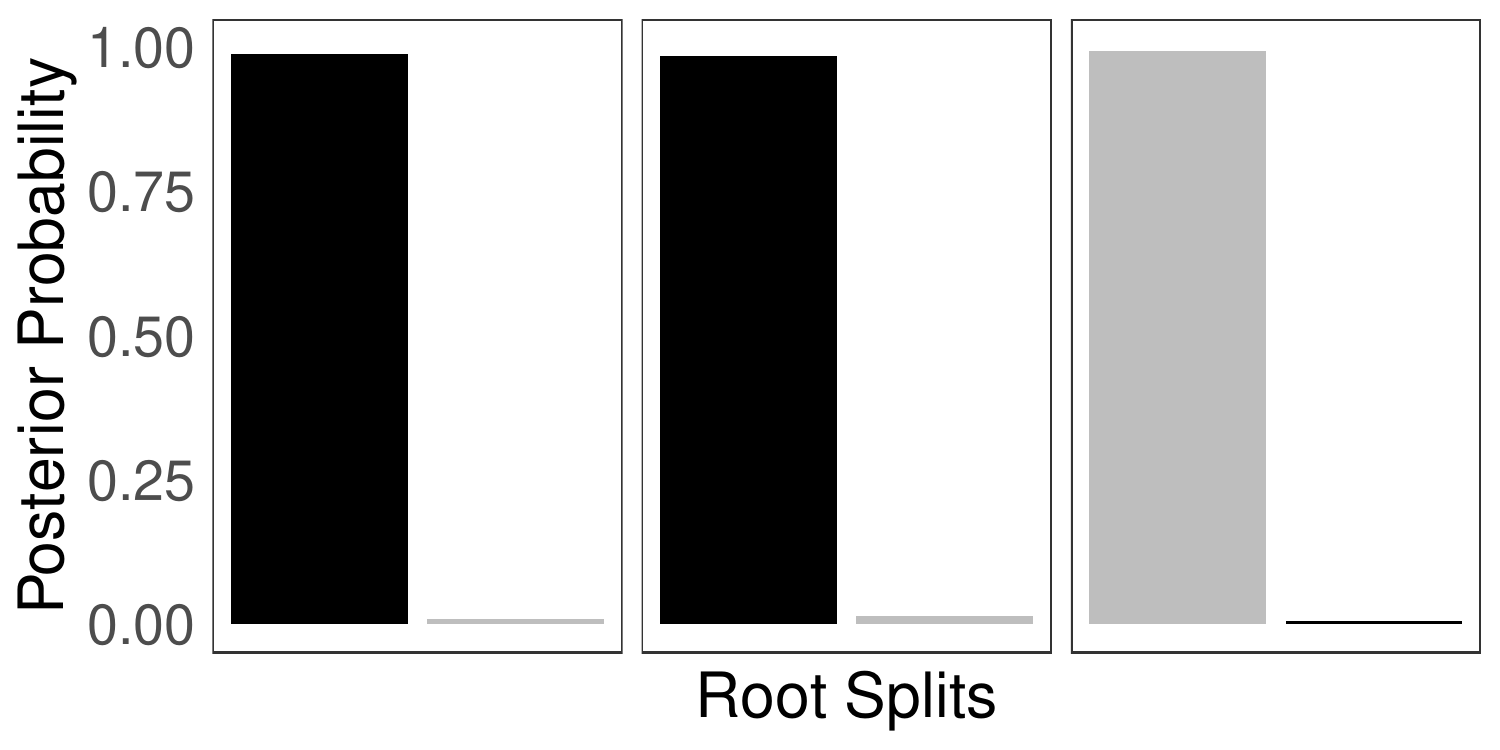}}
\caption{Posterior distribution over roots splits when three data sets are simulated and analysed under the non-homogeneous RY8.8a model and the tree used for simulation is Tree \protect\subref{fig:m6_bl_root} $1$ (balanced, long root branch), \protect\subref{fig:m6_ul_root} $2$ (unbalanced, long internal branch), \protect\subref{fig:m6_bs_root} $3$ (balanced, short root branch), \protect\subref{fig:m6_us_root} $4$ (unbalanced, short internal branch), \protect\subref{fig:m6_bm_root} $5$ (balanced, medium root branch), \protect\subref{fig:m6_um_root} $6$ (unbalanced, medium internal branch). In every plot, bars are arranged in descending order of posterior probability and the correct root split is highlighted in black.}\label{fig:m6_bls_and_topology_root}
\end{figure}

For each of the six trees, three $2000$-site alignments were simulated and analysed under both non-homogeneous Lie Markov models. The posterior distributions over root splits for the RY8.8a model are shown in Figure~\ref{fig:m6_bls_and_topology_root}. In general, root inference is good, with the true root recovered as the posterior mode in most cases. This suggests the posterior is reasonably robust to prior-data conflict concerning the rooted topology and branch lengths. Moreover, for Trees $3$--$6$, whose unrooted trees do not contain any very long edges, the absence of a marked difference between the results for balanced Trees $3$ and $5$ and unbalanced Trees $4$ and $6$ suggest that the prior over rooted topologies imparts little influence over the posterior. It is interesting to note that this was not the case in our earlier work based on a homogeneous, stationary and non-reversible model. However, comparisons between the results for these four trees and Trees $1$ and $2$, which do contain a very long edge, suggest that long branches in the unrooted tree can influence posterior inference of the root position. When the long edge is the root edge (Tree $1$), the posterior is concentrated around the true root position in the analyses of all three alignments; see Figure~\ref{fig:m6_bl_root}. However, when the long edge is not the root edge (Tree $2$), prior-data conflict arises and the true root only has appreciable posterior support in the analysis of one of the three alignments; see Figure~\ref{fig:m6_ul_root}.

The corresponding plots for the non-homogeneous RY5.6b model are shown in Supplementary Figure~S6. In keeping with the results in Section~\ref{subsec:taxa_and_sites} for the larger trees on 12 or 24 taxa, the true root rarely receives particularly appreciable posterior support. In fact, the only cases where the true root was recovered as the posterior mode were the analyses of the three alignments simulated under Tree $1$, where the root edge is a long branch. This further suggests that for alignments of around 2000 sites on a modest number of taxa, the likelihood of a non-homogeneous RY5.6b model does not clearly identify the position of the root.

It is worth noting that for both models and all trees, inference of the unrooted topology was excellent, with the true unrooted topology identified as the posterior mode, with high support, in all cases; see Supplementary Figures~S7 and S8.

\section{\label{sec:application}Application}

To illustrate the benefit of our non-homogeneous model in its facility to bring two sources of information to bear on the rooting problem, we consider an application to a \textit{Drosophila} dataset, taken from \citet*{tarrio}, where most models fail to identify a plausible root position. The alignment contains 2085 nucleotides (sites) from the xanthine dehydrogenase (Xdh) gene of 17 different species of \textit{Drosophila}. \emph{D. saltans}, \emph{D. prosaltans}, \emph{D. neocordata}, \emph{D. emarginata}, \emph{D. sturtevanti and D. subsaltans} form a clade of \emph{saltans}. Three of the species form an outgroup: \emph{D. melanogaster, D. virilis} and \emph{D. pseudoobscura}.  The remaining species form a clade of \emph{willistoni}. Here the term \emph{clade} refers to the subset of taxa obtained by cutting a rooted tree on a branch and selecting only those leaves which are descendants of the split lineage; or, in biological terms, an ancestor and all its descendants. The corresponding concept for unrooted trees is a \emph{split}, which is a bipartition of the taxa into two disjoint sets, induced by cutting a branch.


\subsection{Models}

We compare inferences obtained under six different models:
\begin{description}
\item[$\mathcal{M}_1$:] homogeneous, stationary, reversible GTR model;
\item[$\mathcal{M}_2$:] homogeneous, stationary, non-reversible RY5.6b model;
\item[$\mathcal{M}_3$:] homogeneous, stationary, non-reversible RY8.8a model;
\item[$\mathcal{M}_4$:] non-homogeneous, non-stationary, locally reversible GTR model;
\item[$\mathcal{M}_5$:] non-homogeneous, non-stationary, locally non-reversible RY5.6b model;
\item[$\mathcal{M}_6$:] non-homogeneous, non-stationary, locally non-reversible RY8.8a model.
\end{description}
We note that the likelihood for our baseline model $\mathcal{M}_1$ is invariant to the position of the root and so can only distinguish between unrooted trees. However, as the data set contains an outgroup, we can apply the standard approach of outgroup rooting to polarise the relationships on the unrooted trees with the highest posterior support.

\subsection{Prior specification}
\label{subsec:prior_spec}

For the non-homogeneous Lie Markov models, $\mathcal{M}_5$ and $\mathcal{M}_6$, we use the priors described in Section~\ref{subsec:taxa_and_sites} for the analysis of simulated data. For the remaining models, $\mathcal{M}_i$, $i=1,\ldots,4$, our prior specification is detailed in Section~S3.1 of the Supplementary Materials.

\subsection{MCMC implementation}

For each model, we ran the MCMC algorithm for at least 300K iterations, omitting all but the last 100K as burn-in and thinning the remaining output to retain every 100-th iteration so as to reduce computational overheads.  To rigorously assess convergence and mixing we follow the methods utilised by \citet{heaps}. In brief, for each analysis, we run two chains initialised at different starting states. We then consider standard graphical diagnostics, such as trace and density plots, for the quantitative parameters and assess mixing and convergence in tree space using plots of the cumulative relative frequencies of sampled splits (for model $\mathcal{M}_1$) or clades (for models $\mathcal{M}_2$ -- $\mathcal{M}_6$) over the course of the MCMC run. These checks gave no evidence of any lack of convergence and thinning to every 100-th iterate seemed to produce near-uncorrelated posterior samples.

\subsection{Posterior inference}

\begin{figure}[!ht]
\centering
\subfloat[\label{fig:s_consensus_m1}]{\includegraphics[width=0.45\textwidth]{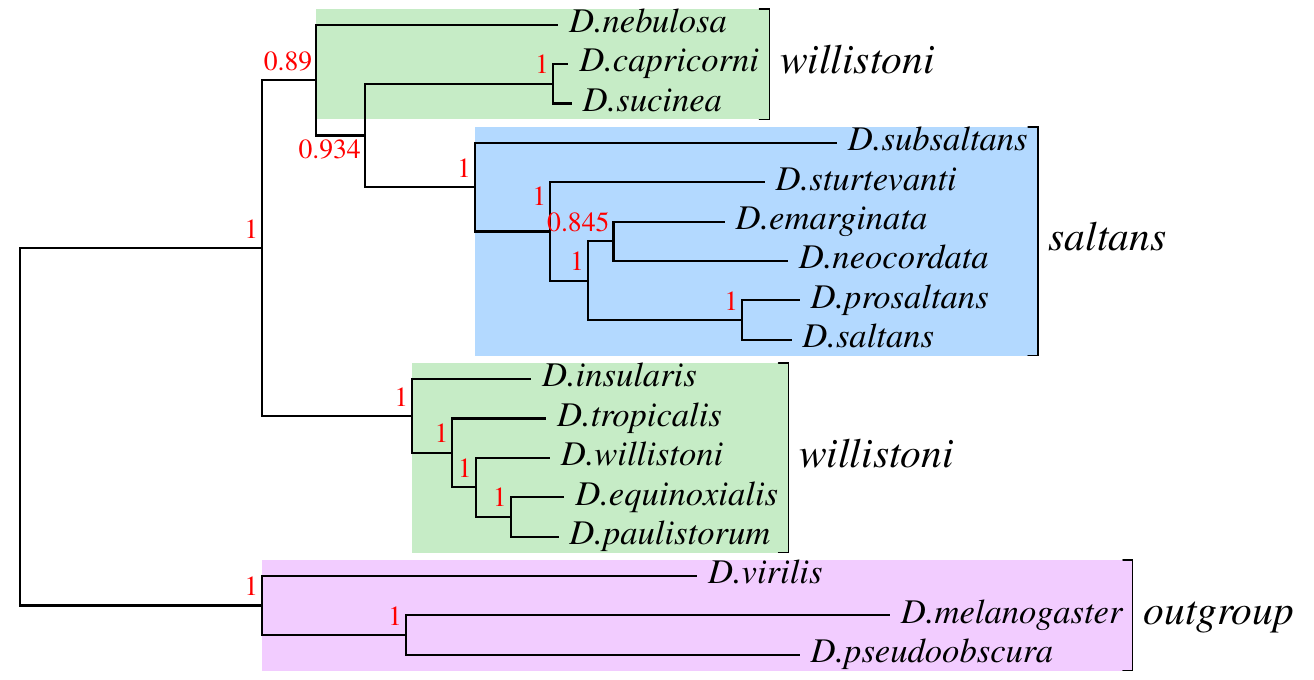}}
\hfill
\subfloat[\label{fig:s_consensus_m2}]{\includegraphics[width=0.45\textwidth]{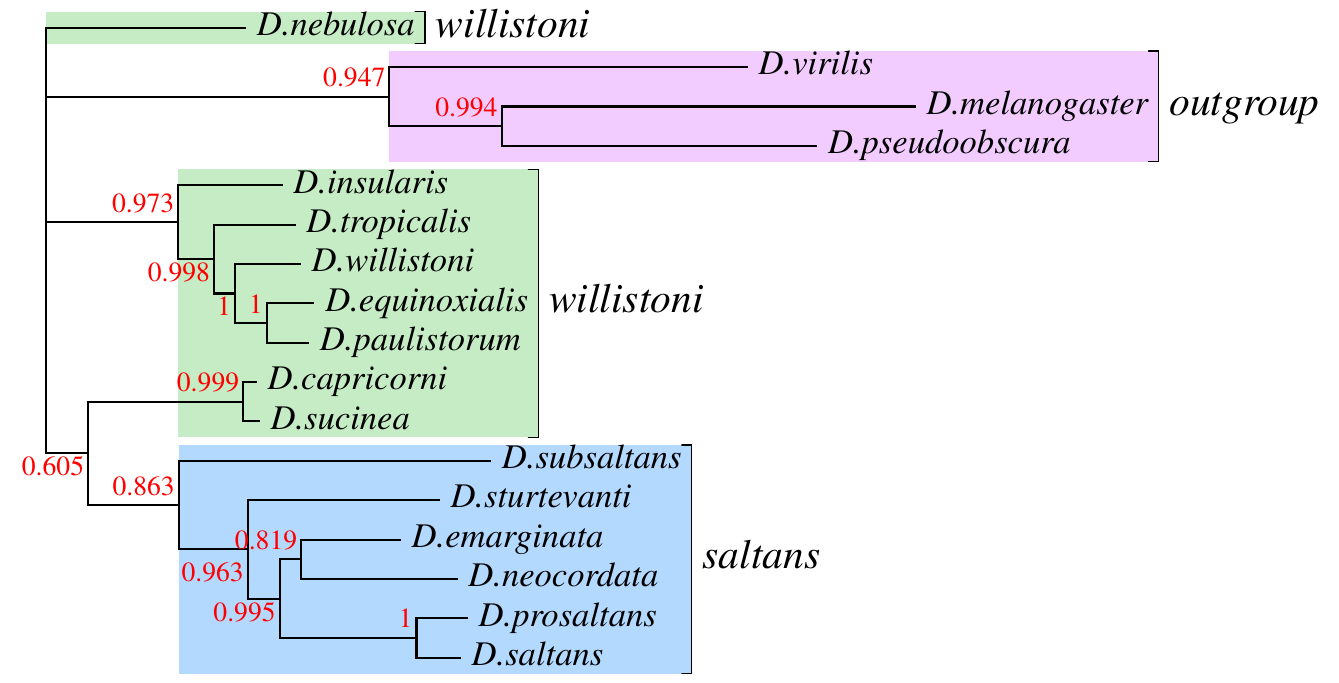}}\\
\subfloat[\label{fig:s_consensus_m3}]{\includegraphics[width=0.45\textwidth]{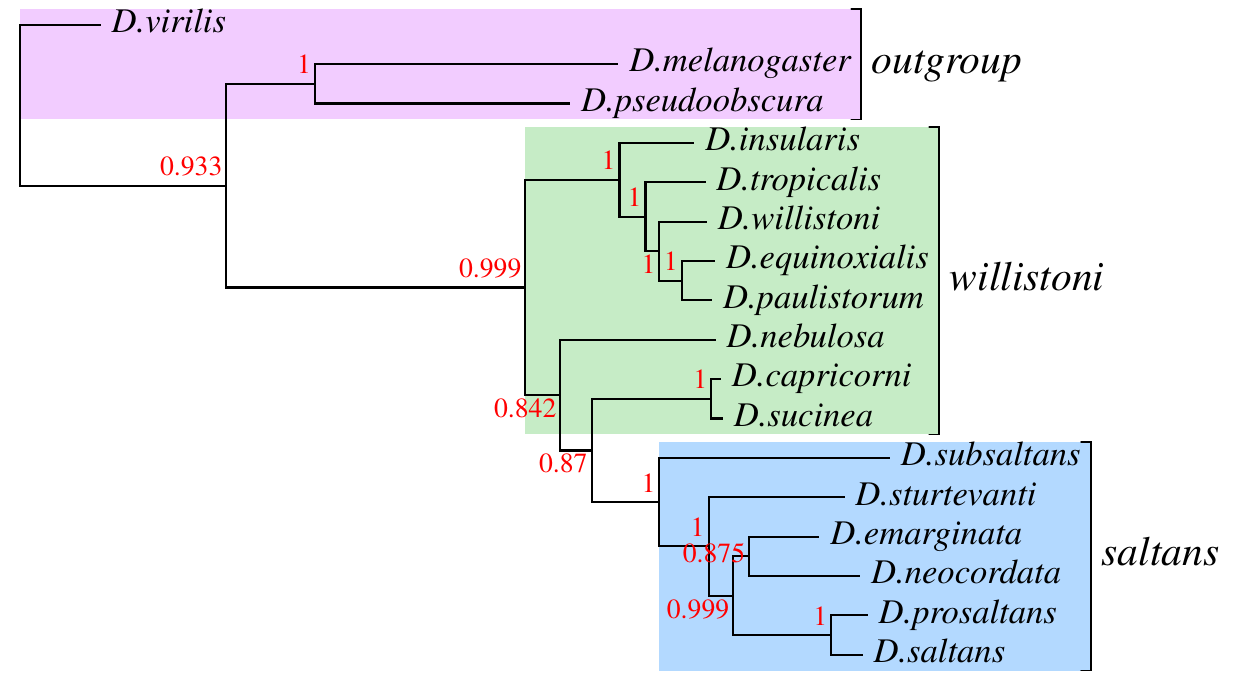}}
\hfill
\subfloat[\label{fig:s_consensus_m4}]{\includegraphics[width=0.45\textwidth]{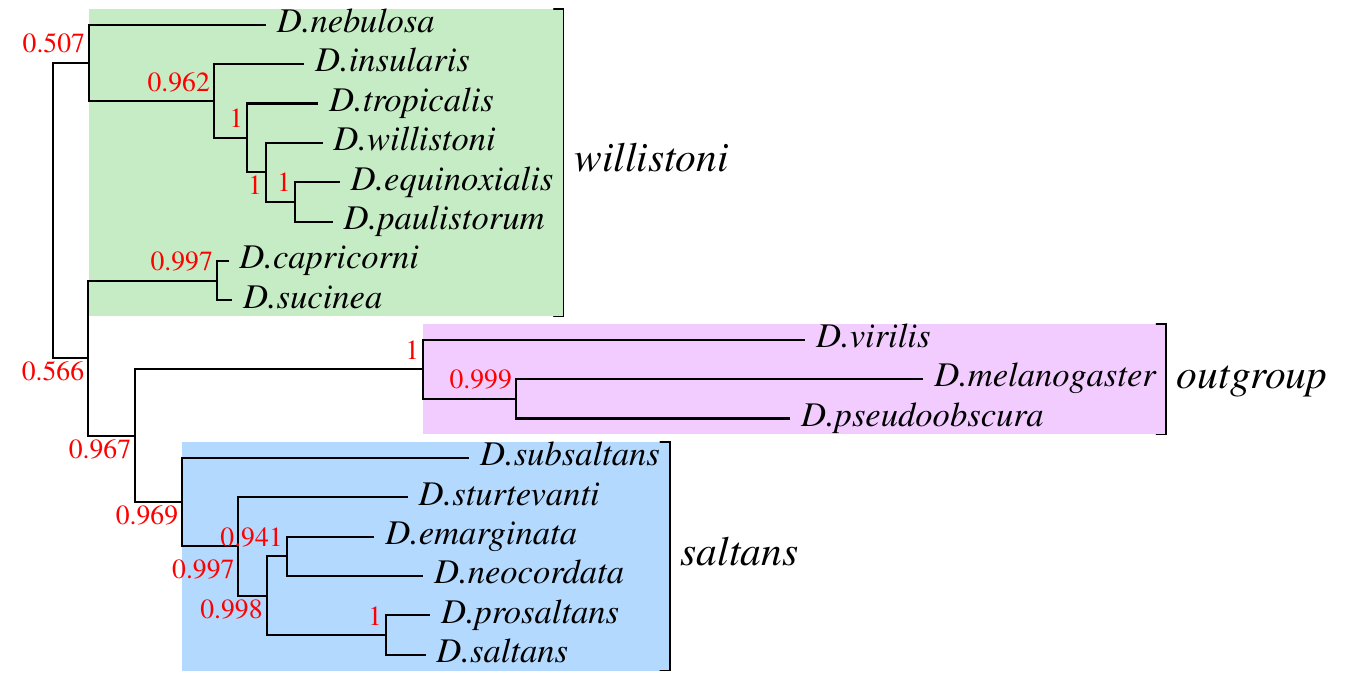}}\\
\subfloat[\label{fig:s_consensus_m5}]{\includegraphics[width=0.45\textwidth]{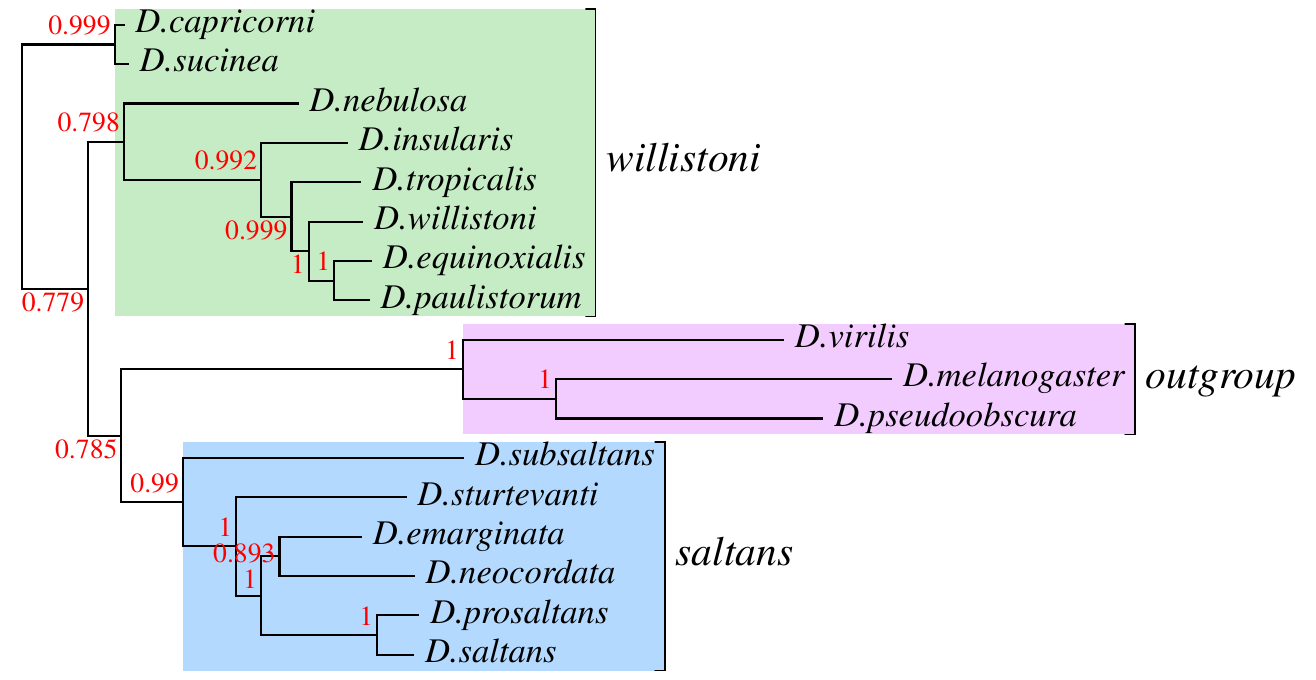}}
\hfill
\subfloat[\label{fig:s_consensus_m6}]{\includegraphics[width=0.45\textwidth]{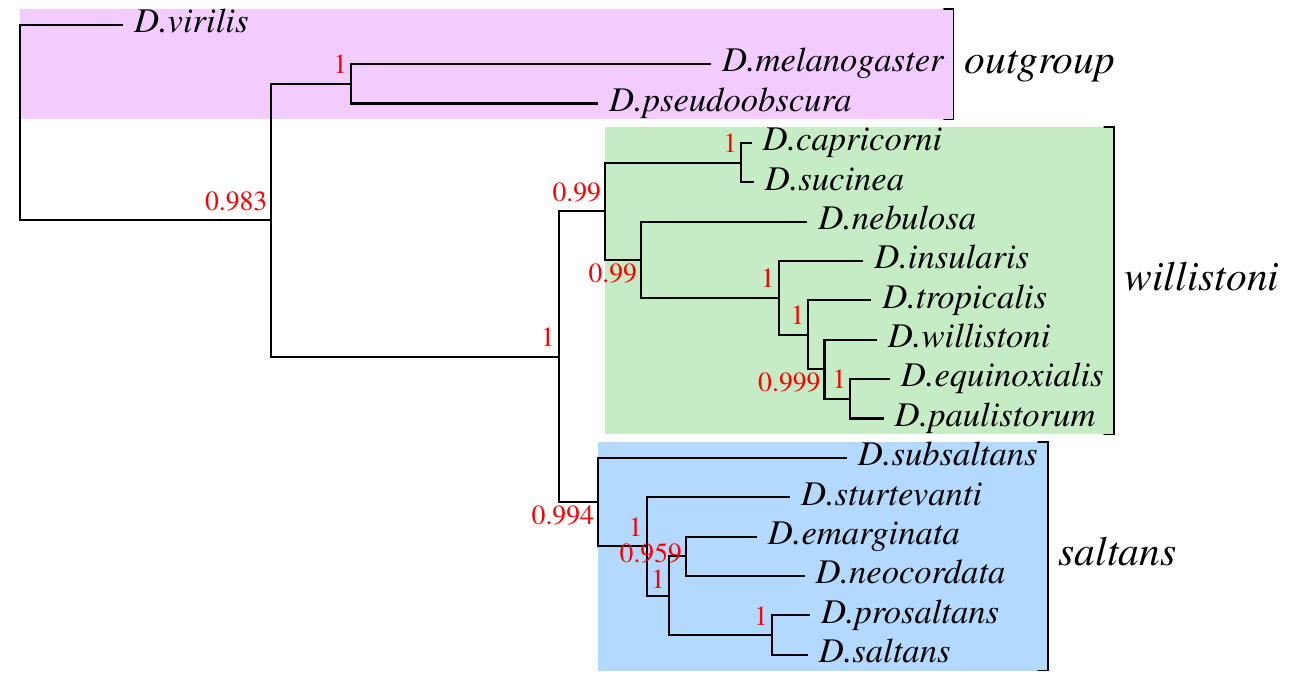}}
\caption{Majority rule consensus trees under the models \protect\subref{fig:s_consensus_m1} $\mathcal{M}_1$ - GTR; \protect\subref{fig:s_consensus_m2} $\mathcal{M}_2$ - RY5.6b; \protect\subref{fig:s_consensus_m3} $\mathcal{M}_3$ - RY8.8a; \protect\subref{fig:s_consensus_m4} $\mathcal{M}_4$ - non-homogeneous GTR; \protect\subref{fig:s_consensus_m5} $\mathcal{M}_5$ - non-homogeneous RY5.6b; \protect\subref{fig:s_consensus_m6} $\mathcal{M}_6$ - non-homogeneous RY8.8a. Numerical labels represent the posterior probability of the associated split (in~\protect\subref{fig:s_consensus_m1}) or clade (in \protect\subref{fig:s_consensus_m2} -- \protect\subref{fig:s_consensus_m6}).}\label{fig:s_consensus}
\end{figure}

In phylogenetic inference, the majority-rule consensus tree is the most widely used summary of the posterior distribution over tree space. As a summary of a sample of trees, it includes only those splits (for unrooted trees) or clades (for rooted trees) which appear in over half of the samples \citep[][]{Bry03}, here representing those with posterior probability greater than 0.5. The consensus trees for the homogeneous and stationary models $\mathcal{M}_1$ -- $\mathcal{M}_3$ are shown in Figures~\ref{fig:s_consensus_m1} -- \ref{fig:s_consensus_m3}, in which numerical labels represent the posterior probability of the associated split ($\mathcal{M}_1$) or clade ($\mathcal{M}_2$, $\mathcal{M}_3$). The majority-rule consensus tree obtained using the GTR model $\mathcal{M}_1$ is unrooted but has been visualised with the root at the midpoint of the branch leading to the outgroup, in accordance with the method of outgroup rooting. We see immediately that the branching structure of the underlying unrooted topology prevents identification of a monophyly (clade) of \emph{willistoni}. Apart from some lack of resolution in the RY5.6b consensus tree, those for the two homogeneous and stationary Lie Markov models, $\mathcal{M}_2$ and $\mathcal{M}_3$, represent the same unrooted topology as the majority-rule tree for the GTR model $\mathcal{M}_1$. As such, they also fail to isolate the \emph{willistoni} as a clade. We note, however, that whilst the RY8.8a consensus tree has a biologically implausible branching structure, its root position, within the outgroup, represents a credible evolutionary hypothesis. Indeed under the RY8.8a model $\mathcal{M}_3$, the posterior probability for a root position within the outgroup or on its parent branch is 0.999, compared to only 0.343 for the simpler RY5.6b model $\mathcal{M}_2$.  

The majority-rule consensus trees for the non-homogeneous and non-stationary models $\mathcal{M}_4$ -- $\mathcal{M}_6$ are shown in Figures~\ref{fig:s_consensus_m4} -- \ref{fig:s_consensus_m6}. All three trees depict the same underlying unrooted topology. This differs from that obtained under the three homogeneous models, and is now biologically plausible, with the \emph{willistoni}, \emph{saltans} and \emph{outgroup} species forming a tripartition, induced by cutting two edges. However, only the non-homogeneous RY8.8a model $\mathcal{M}_6$ identifies a credible root position, with the root on the consensus tree appearing inside the outgroup, and the marginal posterior probability for a root within the outgroup, or on its parent branch, equal to 1.000. In contrast, the roots on the consensus trees for the non-homogeneous GTR and RY5.6b models, $\mathcal{M}_4$ and $\mathcal{M}_5$, split the \emph{willistoni}, whilst the marginal posterior probability for a root within the outgroup, or on its parent branch, is equal to 0.000 in each case. It is interesting to note that under model $\mathcal{M}_6$, the posterior is not only centred on a plausible tree, it is also concentrated in its vicinity, with the posterior for rooted trees notably less diffuse than the distribution obtained under other models. For instance, the rooted topology of the consensus tree, depicted in Figure~\ref{fig:s_consensus_m6}, has posterior probability equal to 0.9235, compared to posterior probabilities of at most 0.6870 for the modes in other cases. This greater concentration of the posterior for the unknowns in $\mathcal{M}_6$ compared with those in $\mathcal{M}_5$ is consistent with the results from the analyses of simulated data in Section~\ref{sec:simulations}.

\subsection{Model comparison}
As is commonly observed in statistical phylogenetics, our phylogenetic inferences are sensitive to the choice of substitution model. One way to arbitrate this inconsistency is through comparisons of the fits of different models; notionally, we have less reason to refute the conclusions of a model which shows a better fit to the data. A natural measure of model uncertainty in the Bayesian framework is the posterior mass function over models, in this case $\Pr(\mathcal{M}_i | y) \propto p(y | \mathcal{M}_i) \Pr(\mathcal{M}_i)$ for $i=1,\ldots,6$, which reduces to $\Pr(\mathcal{M}_i | y) \propto p(y | \mathcal{M}_i)$ in the case of equal prior probabilities, $\Pr(\mathcal{M}_i) = 1 / 6$. For each model $\mathcal{M}_i$, the crucial component is therefore the marginal, or integrated, likelihood $p(y | \mathcal{M}_i)$, given by
\begin{equation*}
p(y | \mathcal{M}_i) = \textstyle\sum_{\tau} \int_{\Theta_i} p(y | \tau, \Theta_i, \mathcal{M}_i) \pi(\tau, \Theta_i | \mathcal{M}_i) d \Theta_i.
\end{equation*}
Here $\Theta_i$ denotes the collection of continuous-valued model parameters, $\vect{\ell}$, $\alpha$ and $\mathcal{Q}_i$, for model $\mathcal{M}_i$.

Numerical calculation of the marginal likelihood is a notoriously difficult computational challenge. This is particularly true in phylogenetics due to the discrete nature of tree space; see \citet{OCML19} for a recent review. Among the various approximations that have been proposed, many techniques are based on importance sampling, with more successful methods typically introducing intermediate (importance) densities to bridge the gap between the prior and posterior. One such method, which is computationally tenable for the complex non-homogeneous models introduced here, is the hybrid estimator of \citet{NR94}, in which the importance density is a mixture between the prior and posterior. Further discussion and full details of the algorithm and can be found in Section~S3.3 of the Supplementary Materials. The log marginal likelihoods for models $\mathcal{M}_1$ -- $\mathcal{M}_6$, approximated using the Newton and Raftery hybrid estimator are displayed in Table~\ref{tab:hybrid}. 

\begin{table}[tb]
\centering
\renewcommand{\arraystretch}{1.25}
\begin{tabular}{cccccccc}\hline
\small
$\mathcal{M}_1$ &$\mathcal{M}_2$ &$\mathcal{M}_3$ &$\mathcal{M}_4$ &$\mathcal{M}_5$ &$\mathcal{M}_6$ \\ \hline
-14719.55 &-14750.98 &-14719.95 &-14531.11 &-14625.97 &\textbf{-14508.22}\\ \hline
\end{tabular}
\caption{\label{tab:hybrid}Log marginal likelihoods for each model approximated using the Newton and Raftery hybrid estimator in which the prior weight in the importance density was set at $\delta = 0.05$. The models are: $\mathcal{M}_1$ - GTR; $\mathcal{M}_2$ - RY5.6b; $\mathcal{M}_3$ - RY8.8a; $\mathcal{M}_4$ - non-homogeneous GTR; $\mathcal{M}_5$ - non-homogeneous RY5.6b; $\mathcal{M}_6$ - non-homogeneous RY8.8a.}
\end{table}

The superior model fit afforded by the three non-homogeneous and non-stationary models is immediately apparent from Table~\ref{tab:hybrid}. For instance, if we perform pairwise comparisons between each homogeneous model and its non-homogeneous counterpart, then the log Bayes factor ranges from 125.01 to 211.73 in favour of the non-homogeneous model. Of the three non-homogeneous models, the non-stationary RY8.8a model seems to give the best fit to the data. This is also the only model whose posterior supported a biologically credible rooted tree. Reasons for its superiority over the simpler RY5.6b variant were discussed from a theoretical and practical perspective in Sections~\ref{subsec:ry88a} and \ref{subsec:taxa_and_sites}, respectively, whilst the improvement over the non-homogeneous GTR model may be attributable to the additional source of root information gained through the non-reversible structure of the RY8.8a rate matrix.

In an attempt to account for the variance of the Newton and Raftery hybrid estimator, we ran ten MCMC chains for each model, initialised at different starting points, and repeated our calculations. The results, shown in Figure~S9 of the Supplementary Materials, suggest a clear ranking of models, such that our conclusions stand even in light of the Monte Carlo error in the marginal likelihood estimates.

\section{\label{sec:discussion}Discussion}

The CTMP that defines standard substitution models of DNA evolution is typically assumed to be reversible and in its stationary distribution. These assumptions are made primarily for mathematical convenience, despite being refutable by experimental evidence and biological theory, and restrictive in generating a root--invariant likelihood. Both issues can be addressed by relaxing one or both simplifying assumptions. Among models in the literature which facilitate root inference, the most biologically credible are those which allow variation in sequence composition over time.

In earlier work we introduced a class of non-homogeneous and non-stationary models with locally reversible structure. Conditional on a given tree, each branch of the underpinning unrooted topology was associated with its own matrix from a class of reversible rate matrices. The distribution at the root of the tree was taken as the stationary distribution on the rooting branch. We have now advanced this idea so that each rate matrix comes from a class of non-reversible Lie Markov models; either RY5.6b or RY8.8a. For both models, we provided a new parameterisation and gave an interpretation. For the homogeneous RY5.6b model, we showed that the additive structure of the rate matrix makes it ill-suited to modelling evolutionary processes where the long-run proportions of each nucleotide are similar and the transition-transversion rate ratio is high. This provides an explanation for the poor fit that is commonly reported for the RY5.6b model. To our knowledge, this has hitherto gone unnoticed in the literature.
 
Our non-homogeneous Lie Markov models have a number of strengths. With fewer parameters than an analogous non-homogeneous general Markov model, they provide a parsimonious way of introducing local non-reversible structure into a non-stationary model. This yields an extra source of information about the root position, whilst retaining computational tractability in model-fitting. Moreover, because Lie Markov models are closed under matrix multiplication, our non-homogeneous extensions are mathematically consistent, meaning the distributions over DNA characters induced by a tree, and all its sub-trees, could have arisen from the same family of non-homogeneous Lie Markov models.

Taking a Bayesian approach to inference, we describe a prior for the branch-specific parameters that encourages borrowing of strength between edges. This has a regularising effect on the posterior distribution. We additionally describe an MCMC scheme for generating samples from the posterior. Through extensive simulation experiments, we demonstrated, empirically, that the root position can be identified from the likelihood of our non-homogeneous models, and that increasing the number of sites in the alignment tends to lead to more accurate and precise inferences of all unknowns. Whilst root inference for the non-homogeneous RY5.6b model was generally poor for larger trees, we showed that root inference under the non-homogeneous RY8.8a model remains strong, even in the face of prior-data conflict arising from an unbalanced rooted topology, though inference can be sensitive to the presence of long branches in the unrooted topology.

We utilised our model and inferential procedures in a biological application concerning a challenging data set of \emph{Drosophila}, in which simpler models typically fail to identify a plausible root position. In this analysis, our non-homogeneous RY8.8a model identified a rooted tree that was biologically credible. We showed that this model had the highest marginal likelihood, indicating better fit to the data.

\section*{Supplementary Materials}
A software implementation can be found through the web-page

\centerline{\url{http://www.mas.ncl.ac.uk/~nseg4/LieMarkov/}}

We also provide a document containing supplementary text and figures.

\section*{Acknowledgements}
This work is supported by the Engineering and Physical Sciences Research Council, Centre for Doctoral Training in Cloud Computing for Big Data (grant number EP/L015358/1). We are grateful to two reviewers and the editor for their helpful comments on an earlier version of this paper.



\bibliography{refs}

\begin{thebibliography}{}

\bibitem[\protect\citeauthoryear{Barry and Hartigan}{Barry and
  Hartigan}{1987}]{barryHartigan}
Barry, D. and J.~A. Hartigan (1987).
\newblock Statistical analysis of hominoid molecular evolution.
\newblock {\em Statistical Science\/}~{\em 2}, 191--210.

\bibitem[\protect\citeauthoryear{Blanquart and Lartillot}{Blanquart and
  Lartillot}{2006}]{blanquart}
Blanquart, S. and N.~Lartillot (2006).
\newblock A {Bayesian} compound stochastic process for modeling nonstationary
  and nonhomogeneous sequence evolution.
\newblock {\em Molecular Biology and Evolution\/}~{\em 23\/}(11), 2058--2071.

\bibitem[\protect\citeauthoryear{Bohlin, Eldholm, Pettersson, Brynildsrud, and
  Snipen}{Bohlin et~al.}{2017}]{BEPBS17}
Bohlin, J., V.~Eldholm, J.~H.~O. Pettersson, O.~Brynildsrud, and L.~Snipen
  (2017).
\newblock The nucleotide composition of microbial genomes indicates
  differential patterns of selection on core and accessory genomes.
\newblock {\em BMC Genomics\/}~{\em 18\/}(1), 151.

\bibitem[\protect\citeauthoryear{Braun and Kimball}{Braun and
  Kimball}{2002}]{braun2002}
Braun, E.~L. and R.~T. Kimball (2002, 07).
\newblock Examining basal avian divergences with mitochondrial sequences: Model
  complexity, taxon sampling, and sequence length.
\newblock {\em Systematic Biology\/}~{\em 51\/}(4), 614--625.

\bibitem[\protect\citeauthoryear{Bryant}{Bryant}{2003}]{Bry03}
Bryant, D. (2003).
\newblock A classification of consensus methods for phylogenies.
\newblock In M.~Janowitz, F.-J. Lapointe, F.~R. McMorris, B.~Mirkin, and F.~S.
  Roberts (Eds.), {\em Bioconsensus, DIMACS Series}, Providence, Rhode Island,
  pp.\  163--184. American Mathematical Society.

\bibitem[\protect\citeauthoryear{Cherlin, Heaps, Nye, Boys, Williams, and
  Embley}{Cherlin et~al.}{2017}]{cherlin17}
Cherlin, S., S.~Heaps, T.~Nye, R.~Boys, T.~Williams, and T.~Embley (2017, 11).
\newblock The effect of non-reversibility on inferring rooted phylogenies.
\newblock {\em Molecular Biology and Evolution\/}.

\bibitem[\protect\citeauthoryear{Cox, Foster, Hirt, Harris, and Embley}{Cox
  et~al.}{2008}]{cox}
Cox, C.~J., P.~G. Foster, R.~P. Hirt, S.~Harris, and T.~M. Embley (2008).
\newblock The archaebacterial origin of eukaryotes.
\newblock {\em Proceedings of the National Academy of Sciences\/}~{\em
  105\/}(51), 20356--20361.

\bibitem[\protect\citeauthoryear{Dutheil and Boussau}{Dutheil and
  Boussau}{2008}]{dutheilBoussau08}
Dutheil, J. and B.~Boussau (2008).
\newblock Non-homogeneous models of sequence evolution in the {Bio++} suite of
  libraries and programs.
\newblock {\em BMC Evolutionary Biology\/}~{\em 28}, 255.

\bibitem[\protect\citeauthoryear{Felsenstein}{Felsenstein}{1973}]{felsenstein}
Felsenstein, J. (1973).
\newblock Maximum likelihood and minimum-steps methods for estimating
  evolutionary trees from data on discrete characters.
\newblock {\em Systematic Zoology\/}~{\em 22}, 240--249.

\bibitem[\protect\citeauthoryear{Fern\'{a}ndez-S\'{a}nchez, Sumner, Jarvis, and
  Woodhams}{Fern\'{a}ndez-S\'{a}nchez et~al.}{2016}]{fernsand2015}
Fern\'{a}ndez-S\'{a}nchez, J., J.~G. Sumner, P.~D. Jarvis, and M.~D. Woodhams
  (2016).
\newblock Lie {M}arkov models with purine/pyrimidine symmetry.
\newblock {\em Journal of Mathematical Biology\/}~{\em 70\/}(4), 855--891.

\bibitem[\protect\citeauthoryear{Foster}{Foster}{2004}]{foster}
Foster, P.~G. (2004).
\newblock Modeling compositional heterogeneity.
\newblock {\em Systematic Biology\/}~{\em 53\/}(3), 485.

\bibitem[\protect\citeauthoryear{Hasegawa, Kishino, and Yano}{Hasegawa
  et~al.}{1985}]{hasegawa}
Hasegawa, M., H.~Kishino, and T.-a. Yano (1985, Oct).
\newblock Dating of the human-ape splitting by a molecular clock of
  mitochondrial {DNA}.
\newblock {\em Journal of Molecular Evolution\/}~{\em 22\/}(2), 160--174.

\bibitem[\protect\citeauthoryear{Heaps, Nye, Boys, Williams, and Embley}{Heaps
  et~al.}{2014}]{heaps}
Heaps, S.~E., T.~M.~W. Nye, R.~J. Boys, T.~A. Williams, and T.~M. Embley
  (2014).
\newblock Bayesian modelling of compositional heterogeneity in molecular
  phylogenetics.
\newblock {\em Statistical Applications in Genetics and Molecular
  Biology\/}~{\em 1}, 1--21.

\bibitem[\protect\citeauthoryear{Hodgkinson and Eyre-{W}alker}{Hodgkinson and
  Eyre-{W}alker}{2011}]{HE11}
Hodgkinson, A. and A.~Eyre-{W}alker (2011).
\newblock Variation in the mutation rate across mammalian genomes.
\newblock {\em Nature Reviews Genetics\/}~{\em 12\/}(3), 756--766.

\bibitem[\protect\citeauthoryear{Huelsenbeck, Bollback, and Levine}{Huelsenbeck
  et~al.}{2002}]{huelsenbeck2002}
Huelsenbeck, J.~P., J.~P. Bollback, and A.~M. Levine (2002).
\newblock Inferring the root of a phylogenetic tree.
\newblock {\em Systematic Biology\/}~{\em 51\/}(1), 32--43.

\bibitem[\protect\citeauthoryear{Kaehler}{Kaehler}{2017}]{kaehler2017}
Kaehler, B.~D. (2017).
\newblock Full reconstruction of non-stationary strand-symmetric models on
  rooted phylogenies.
\newblock {\em Journal of Theoretical Biology\/}~{\em 420}, 144--151.

\bibitem[\protect\citeauthoryear{Lartillot, Blanquart, and Lepage}{Lartillot
  et~al.}{2004}]{PhyloBayesManual}
Lartillot, N., S.~Blanquart, and T.~Lepage (2004).
\newblock {\em Phylo{B}ayes 3.3: A {B}ayesian software for phylogenetic
  reconstruction and molecular dating using mixture models}.

\bibitem[\protect\citeauthoryear{Li, Shao, Song, Song, Jiang, Li, and Cai}{Li
  et~al.}{2015}]{li2015}
Li, H., R.~Shao, N.~Song, F.~Song, P.~Jiang, Z.~Li, and W.~Cai (2015).
\newblock Higher-level phylogeny of paraneopteran insects inferred from
  mitochondrial genome sequences.
\newblock {\em Scientific Reports\/}~{\em 5}, 8527.

\bibitem[\protect\citeauthoryear{Lind and Andersson}{Lind and
  Andersson}{2008}]{LA08}
Lind, P.~A. and D.~I. Andersson (2008).
\newblock Whole--genome mutational biases in bacteria.
\newblock {\em Proceedings of the National Academy of Sciences of the United
  States of America\/}~{\em 105\/}(46), 17878--17883.

\bibitem[\protect\citeauthoryear{Morgan, Foster, Webb, Pisani, and
  {O'Connell}}{Morgan et~al.}{2013}]{morgan2013}
Morgan, C.~C., P.~G. Foster, A.~E. Webb, D.~Pisani, and J.~O.~M. M.~J.
  {O'Connell} (2013).
\newblock Heterogeneous models place the root of the placental mammal
  phylogeny.
\newblock {\em Molecular Biology and Evolution\/}~{\em 30\/}(9), 2145--2156.

\bibitem[\protect\citeauthoryear{Newton and Raftery}{Newton and
  Raftery}{1994}]{NR94}
Newton, M.~A. and A.~E. Raftery (1994).
\newblock Approximate {Bayesian} inference by the weighted likelihood bootstrap
  (with discussion).
\newblock {\em Journal of the Royal Statistical Society: Series B\/}~{\em 56},
  3--48.

\bibitem[\protect\citeauthoryear{Oaks, Cobb, Minin, and Leach\'{e}}{Oaks
  et~al.}{2019}]{OCML19}
Oaks, J.~R., K.~A. Cobb, V.~N. Minin, and A.~D. Leach\'{e} (2019).
\newblock Marginal likelihoods in phylogenetics: A review of methods and
  applications.
\newblock {\em Systematic Biology\/}.

\bibitem[\protect\citeauthoryear{Phillips, McLenachan, Down, Gibb, and
  Penny}{Phillips et~al.}{2006}]{phillips2006}
Phillips, M.~J., P.~A. McLenachan, C.~Down, G.~C. Gibb, and D.~Penny (2006,
  02).
\newblock {Combined Mitochondrial and Nuclear DNA Sequences Resolve the
  Interrelations of the Major Australasian Marsupial Radiations}.
\newblock {\em Systematic Biology\/}~{\em 55\/}(1), 122--137.

\bibitem[\protect\citeauthoryear{Rosenberg, Subramanian, and Kumar}{Rosenberg
  et~al.}{2003}]{RSK03}
Rosenberg, M.~S., S.~Subramanian, and S.~Kumar (2003).
\newblock Patterns of transitional mutation biases within and among mammalian
  genomes.
\newblock {\em Molecular Biology and Evolution\/}~{\em 20\/}(6), 988--993.

\bibitem[\protect\citeauthoryear{Squartini and Arndt}{Squartini and
  Arndt}{2008}]{squartini}
Squartini, F. and P.~F. Arndt (2008).
\newblock Quantifying the stationarity and time reversibility of the nucleotide
  substitution process.
\newblock {\em Molecular Biology and Evolution\/}~{\em 25}, 2525--2535.

\bibitem[\protect\citeauthoryear{Sumner, Fern{\'a}ndez-S{\'a}nchez, and
  Jarvis}{Sumner et~al.}{2012}]{sumner2012a}
Sumner, J.~G., J.~Fern{\'a}ndez-S{\'a}nchez, and P.~D. Jarvis (2012).
\newblock Lie {Markov} models.
\newblock {\em Journal of Theoretical Biology\/}~{\em 298}, 16 -- 31.

\bibitem[\protect\citeauthoryear{Tarr{\'i}o, Rodr{\'i}guez-Trelles, and
  Ayalaa}{Tarr{\'i}o et~al.}{2000}]{tarrio}
Tarr{\'i}o, R., F.~Rodr{\'i}guez-Trelles, and F.~J. Ayalaa (2000).
\newblock Tree rooting with outgroups when they differ in their nucleotide
  composition from the ingroup: The {Drosophila} saltans and willistoni groups,
  a case study.
\newblock {\em Molecular Phylogenetics and Evolution\/}~{\em 16\/}(3), 344 --
  349.

\bibitem[\protect\citeauthoryear{Tavar{\'e}}{Tavar{\'e}}{1986}]{tavare}
Tavar{\'e}, S. (1986).
\newblock Some probabilistic and statistical problems in the analysis of {DNA}
  sequences.
\newblock {\em Lectures on Mathematics in the Life Sciences\/}~{\em 17},
  57--86.

\bibitem[\protect\citeauthoryear{Williams, Heaps, Cherlin, Nye, Boys, and
  Embley}{Williams et~al.}{2015}]{williams15}
Williams, T.~A., S.~E. Heaps, S.~Cherlin, T.~M.~W. Nye, R.~J. Boys, and T.~M.
  Embley (2015).
\newblock New substitution models for rooting phylogenetic trees.
\newblock {\em Philosophical Transactions of the Royal Society B:
  Biological~Sciences\/}~{\em 370\/}(1678).

\bibitem[\protect\citeauthoryear{Woodhams, Fern{\'a}ndez-S{\'a}nchez, and
  Sumner}{Woodhams et~al.}{2015}]{woodhams}
Woodhams, M.~D., J.~Fern{\'a}ndez-S{\'a}nchez, and J.~G. Sumner (2015).
\newblock A new hierarchy of phylogenetic models consistent with heterogeneous
  substitution rates.
\newblock {\em Systematic Biology\/}~{\em 64}, 638--650.

\bibitem[\protect\citeauthoryear{Yang}{Yang}{1994}]{yang1994}
Yang, Z. (1994, 10).
\newblock Maximum likelihood phylogenetic estimation from {DNA} sequences with
  variable rates over sites: Approximate methods.
\newblock ~{\em 39}, 306--14.

\bibitem[\protect\citeauthoryear{Yang and Roberts}{Yang and
  Roberts}{1995}]{yang1995}
Yang, Z. and D.~Roberts (1995).
\newblock On the use of nucleic acid sequences to infer early branchings in the
  tree of life.
\newblock {\em Molecular Biology and Evolution\/}~{\em 12\/}(3), 451--458.

\bibitem[\protect\citeauthoryear{Zwickl and Holder}{Zwickl and
  Holder}{2004}]{ZH04}
Zwickl, D.~J. and M.~T. Holder (2004).
\newblock Model parameterization, prior distributions, and the general
  time-reversible model in {B}ayesian phylogenetics.
\newblock {\em Systematic Biology\/}~{\em 53\/}(6), 877--888.

\end{thebibliography}


\end{document}